\documentclass{acta}

\usepackage{amsmath}
\usepackage{amssymb}
\usepackage{rotating}
\usepackage{natbib}
\usepackage[hidelinks]{hyperref} 
\usepackage{pdflscape}   
\usepackage{changepage} 
\usepackage{threeparttable}

\begin{document}

\begin{Titlepage}

\Title{Bright Long Secondary Period Stars for Follow-up Observations}

\Author{
P.~~I~w~a~n~e~k$^1$,~~
D.\,M.~~S~k~o~w~r~o~n$^1$,~~
G.~~P~o~j~m~a~\'n~s~k~i$^1$~~and~~
I.~~S~o~s~z~y~\'n~s~k~i$^1$~~
}
{$^1$Astronomical Observatory, University of Warsaw, Al. Ujazdowskie 4, 00-478 Warsaw, Poland\\
e-mail: piwanek@astrouw.edu.pl}

\Received{MM DD, YYYY}

\end{Titlepage}

\Abstract{Long secondary period (LSP) variable stars are a subclass of long-period variables (LPV) that exhibit additional long-term variability alongside pulsations. Despite being observed in over 30\% of LPVs, the reason behind the LSP phenomenon is still debated. The most favoured explanation, supported by recent growing evidence, is binarity, where the pulsating giant star has a substellar-mass companion.
To further test this hypothesis, it is important to identify bright LSP variables, for which high-quality spectroscopic and interferometric observations can be obtained more easily. Motivated by the absence of a catalog of bright nearby LSPs, we searched the All Sky Automated Survey (ASAS) data in the \textit{V}-band magnitude range 5.5-14 mag, and for declinations $< +28^\circ$. The resulting catalog contains 23 LSPs, 13 of which are new discoveries. We compare our catalog with the LSP lists available in the literature.
}
{stars: AGB and post-AGB, stars: late-type, stars: variables: general, astronomical databases:catalogs}

\section{Introduction} \label{sec:introduction}

Long-period variables (LPVs) are late-type evolved stars located on the asymptotic giant branch (AGB) or the upper red giant branch (RGB). They are characterized by long-period radial and often non-radial pulsations in their extended envelopes \citep{wood2015, trabucchi2017}. In addition to pulsational variability, nearly all LPVs exhibit irregular light variations, mostly caused by material ejected through strong stellar winds at this evolutionary stage. However, there exists yet another component of variability -- the so-called long secondary period (LSP) -- whose physical origin remains unknown to this day.

LSPs represent one of the most intriguing and least understood forms of variability, observed in more than one-third LPVs. The first mention of this slow, periodic brightness modulation dates back nearly a century to the work of \citet{oconnell1933}, who noted a correlation between the short (pulsation) and long (secondary) periods -- with the latter being approximately ten times longer than the former. In the following decades, two additional studies based on larger stellar samples confirmed the existence of this phenomenon \citep{payne1954, houk1963}. Despite long-standing observational and theoretical effort, the physical mechanism responsible for LSP variability remains elusive.

During recent years, several hypotheses have been proposed to explain the LSP phenomenon. Some interpretations suggested non-radial pulsations and oscillatory convective modes
\citep{wood2000, saio2015, takayama2020}. However, although such mechanisms can reproduce either the observed periods or amplitudes, they fail to account for the characteristic light curve morphology of stars exhibiting LSP variability. At present, the most widely discussed explanation involves binarity, in which the red giant is accompanied by a low-mass companion, possibly of substellar nature, producing brightness modulation through orbital motion, ellipsoidal variability, or variable obscuration by circumstellar dust \citep{soszynski2007, nicholls2010, soszynski2014, takayama2015, pawlak2024}. A particularly comprehensive and insightful review of the hypotheses proposed to explain the LSP phenomenon was presented by \citet{goldberg2024}.

Recent studies provide increasing support for the hypothesis that LSPs are linked to binarity. \citet{soszynski2021} combined the \textit{Optical Gravitational Lensing Experiment} \citep[OGLE;][]{udalski2015} photometry for LSP stars with mid-infrared (mid-IR) data from the \textit{Near-Earth Object WISE Reactivation Mission} \citep[NEOWISE-R;][]{mainzer2014}.
The authors found that many LSP stars exhibit secondary eclipses visible only at infrared wavelengths, concluding that such behavior is consistent with a dusty cloud orbiting the red giant together with a~companion. This scenario has been supported by recent hydrodynamical modeling \citep{decin2025}. \citet{danilovich2025} used high-resolution ALMA continuum imaging of AGB stars and showed that many of their asymmetric dust morphologies can be naturally explained by the presence of a companion that enhances dust formation in the stellar wind regions. \citet{macleod2025} placed tight constraints on a possible companion to Betelgeuse ($\alpha$ Ori), a canonical bright red supergiant showing LSP variability, and showed that the LSP is consistent across both radial velocity and astrometric datasets, indicating the presence of a low-mass companion. Together, these results strengthen the view that LSP phenomena may arise from red giants orbited by faint, substellar companions, possibly surrounded by -- and dragging along -- a dusty tail.

The vast majority of known LSP variables have been discovered by large-scale sky surveys, such as OGLE \citep{soszynski2009, soszynski2011} and the \textit{MAssive Compact Halo Objects}  survey \citep[MACHO;][]{alcock1992, wood1999} which primarily target the Magellanic Clouds and the Galactic bulge. Consequently, the majority of discovered LSPs are relatively faint and distant, limiting the feasibility of detailed spectroscopic and interferometric follow-up with contemporary instruments. In contrast, bright nearby LSPs offer an invaluable opportunity to investigate this phenomenon in detail using a wide range of methods, including high-resolution spectroscopy, radial velocity monitoring, and interferometric imaging.

Despite their importance, no homogeneous catalog of bright, nearby LSP stars currently exists. Some bright candidates have been reported by \citet{percy2020, percy2022, percy2023, percyshenoy2023, percy2025}, but the available samples typically lack precise and uniform photometric characterization. The \textit{All Sky Automated Surver} \citep[ASAS;][]{pojmanski1997} provides an excellent database to search for such objects across nearly the entire sky. ASAS delivers long-term (the longest ones have 19 years of observations), well-sampled light curves for stars brighter than $V=14$ mag, making it particularly suitable for detecting LSP variability in bright, nearby giants.

In this paper, we present a systematic search for bright stars exhibiting clear LSP variability using ASAS photometric data. Our primary goal is to construct a~well-defined catalog of such objects suitable for high-quality multi-method follow-up observations. The paper is organized as follows. Section \ref{sec:data} describes the ASAS instrumentation, observational strategy, and data reduction procedures. In Section~\ref{sec:selection} we outline the selection and classification methods, while Section \ref{sec:collection} presents the resulting catalog of $23$ LSPs along with their observational parameters, light curves and cross-identifications with external catalogs. Finally, Section \ref{sec:conclusions} summarizes our findings.

\section{Observations} \label{sec:data}

The \textit{All Sky Automated Survey} (ASAS) is a long-term project
dedicated to detecting and monitoring the variability of bright stars. 
Over its 25-year history, it has used two observing sites:
Las Campanas Observatory in Chile (operated by the Carnegie
Institution for Science) and Las Cumbres Observatory in Haleakala
(Maui, Hawaii, USA). The equipment used (CCD cameras and lenses), as well as
the data collection and reduction process, has changed several times.

Data for the southern region (up to declination $+28^\circ$) were initially collected 
using Apogee AP-10 2Kx2K CCD cameras, \textit{V} and \textit{I} filters and 200mm f/2.8 
telephoto lenses. Until 2003, data were collected in single 180-second 
exposures, which resulted in images of stars with $V < 8.5$ mag being saturated and 
making standard photometry useless.

Since 2003, data have been collected in triple exposures of 60 s 
each, and special attention has been paid to saturated observations 
during data analysis to make them useful. This has pushed the saturation 
limit to $V \sim 7.5$ mag, and photometry of even much brighter stars is partially 
useful (although it is characterized by high noise $\gtrsim 0.1$ mag and zero-point 
uncertainty).

In 2010, the cameras were replaced with FLI ProLine 16000 devices equipped 
with 4Kx4K CCDs, and lenses with larger aperture (200mm f/2.0) were installed.
At the same time, the observation scheme was changed so that two exposures 
were made, one 180 s and the other 18 s, which allowed a further 
shift of the saturation limit to about $5.5$-$6$ mag.

At the northern ASAS observing site, from 2006 to 2017, Apogee AP-10 cameras and 200mm/2.0 lenses were used, and single exposures of 180 s each were taken. This means 
that for stars brighter than $V \sim 8.3$ mag, the measurements are significantly 
affected by saturation, and we decided not to use them in this work, as our magnitude limit has been set to $V<8$~mag.

Details about the instruments and the data acquisition and reduction 
pipeline can be found in earlier ASAS publications, e.g. \citet{pojmanski1997, pojmanski2001, pojmanski2002}.

\section{Selection and Classification of LSP} \label{sec:selection}

As a starting point, we used the Tycho-2 Catalog \citep{hog2000}, to select a~sample of bright and red objects, applying the criteria \textit{V}$<8$ mag and $B-V > 0.5 $ mag. Due to the ASAS North saturation limit (see Section \ref{sec:data} for more details), we selected objects with declination less than $28^\circ$. 

This resulted in a sample of $16\;381$ targets. For all of these objects, we then extracted light curves from the ASAS databases \citep{pojmanski1997} yielding a total of $27\;632$ light curves in the \textit{V}-band. The number of light curves exceeds the number of targets because some stars are observed by both southern and northern ASAS telescopes.

ASAS uses aperture photometry as this method has proven more reliable than others for wide field, subsampled CCD images. It also provides useful, albeit lower-quality, photometry of saturated stars \citep{pojmanski1997}. Each ASAS frame is analyzed using five different aperture sizes: 2, 3, 4, 5 and 6 pixels in diameter. For each star, we selected the aperture that yielded the smallest dispersion in the measured magnitudes. Additionally, every ASAS frame is assigned a quality grade from A~(best) to D (worst). For our analysis, we retained only the A-grade measurements.

For the selected light curves, we determined the periods using the standard discrete Fourier transform modified for unevenly spaced data \citep{kurtz1985} as implemented in the \texttt{FNPEAKS}\footnote{\url{http://helas.astro.uni.wroc.pl/deliverables.php?active=fnpeaks}} code by Z. Ko\l aczkowski, W. Hebisch, and G. Kopacki. We searched the frequency space from $0.0005$ to $0.1$ day$^{-1}$ (corresponding to periods from $2000$ to $10$ days) with a frequency resolution of $10^{-5}$ day$^{-1}$, and selected the period corresponding to the highest signal-to-noise ratio. With the light curves and periods determined, we proceeded to the most time-consuming stage of the analysis, i.e, the visual inspection and classification of the variability.

In the first step of visual inspection, we examined by eye each of the $27\;632$ light curves both in their raw (unfolded) form and in their phase-folded form according to the derived period. At this stage, we divided the objects into two categories: non-LSP and probable LSP candidates. The latter group contained $101$ objects. During this step we also removed obvious outlying measurements from the light curves.

In the second step, we substracted the primary period (hereafter $P_L$, corresponding to the LSP) from each of these $101$ light curves and again searched the frequency space from $0.001$ to $0.9$ day$^{-1}$ (i.e., periods from $1000$ days to $\sim 1$~day) with the same frequency resolution. We selected the period corresponding to the highest signal-to-noise ratio and adopted it as the pulsation period (hereafter $P_S$). During this step, we focused primarily on the shape of the variability caused by the LSP, requiring in particular that the LSP amplitude be greater than the amplitude of the stellar pulsation (to avoid confusing LSP with other types of red giant variability, e.g., semi-regular or ellipsoidal variations, dust obscuration). After carefully re-inspecting all $101$ candidates, we classified 23 objects as exhibiting clear and robust signatures of the LSP variability.

For these 23 objects, we fine-tuned the periods $P_L$ and $P_S$ using the \texttt{TATRY} code, which implements the multiharmonic analysis of variance algorithm \citep{schwarzenberg1996}. We did not apply the \texttt{TATRY} code at the initial stage of the analysis, as this method is significantly more computationally demanding than \texttt{FNPEAKS}.
During this step, we first fine-tuned the $P_L$ periods, subtracted them from the light curves, then determined the $P_S$ periods, and finally fine-tuned the $P_S$ values.

\section{Collection} \label{sec:collection}

We present a collection of $23$ bright stars with clear LSP variability identified in the ASAS data. Thirteen of these objects are new discoveries, while ten of them have previously been recognized as LSP variables, i.e., BM~Eri, S~Lep, NSV~16816, GO~Vel, MN~Vel, Y~Hya, IQ~Her, V0988~Oph, U~Del, CI~Phe. Table \ref{tab:lsp} summarizes the properties of all stars in the sample, including the ASAS identifier, equatorial coordinates (R.A. and Decl.), mean \textit{V}-band magnitude (\textit{V}), \textit{V}-band amplitude ($A_V$), long secondary period ($P_L$), and pulsation period ($P_S$).

The mean \textit{V}-band magnitude ($V$) was computed as the median (50th percentile) of the flux distribution, and the amplitude $A_V$ was estimated as the difference between the 95th and 5th percentiles of the flux distribution. The measurements were first converted from magnitudes to flux units, the percentiles were then computed in the flux space, and the resulting median flux and amplitude were finally converted back to magnitudes.

\begin{figure}[h]
\includegraphics[width=1\textwidth]{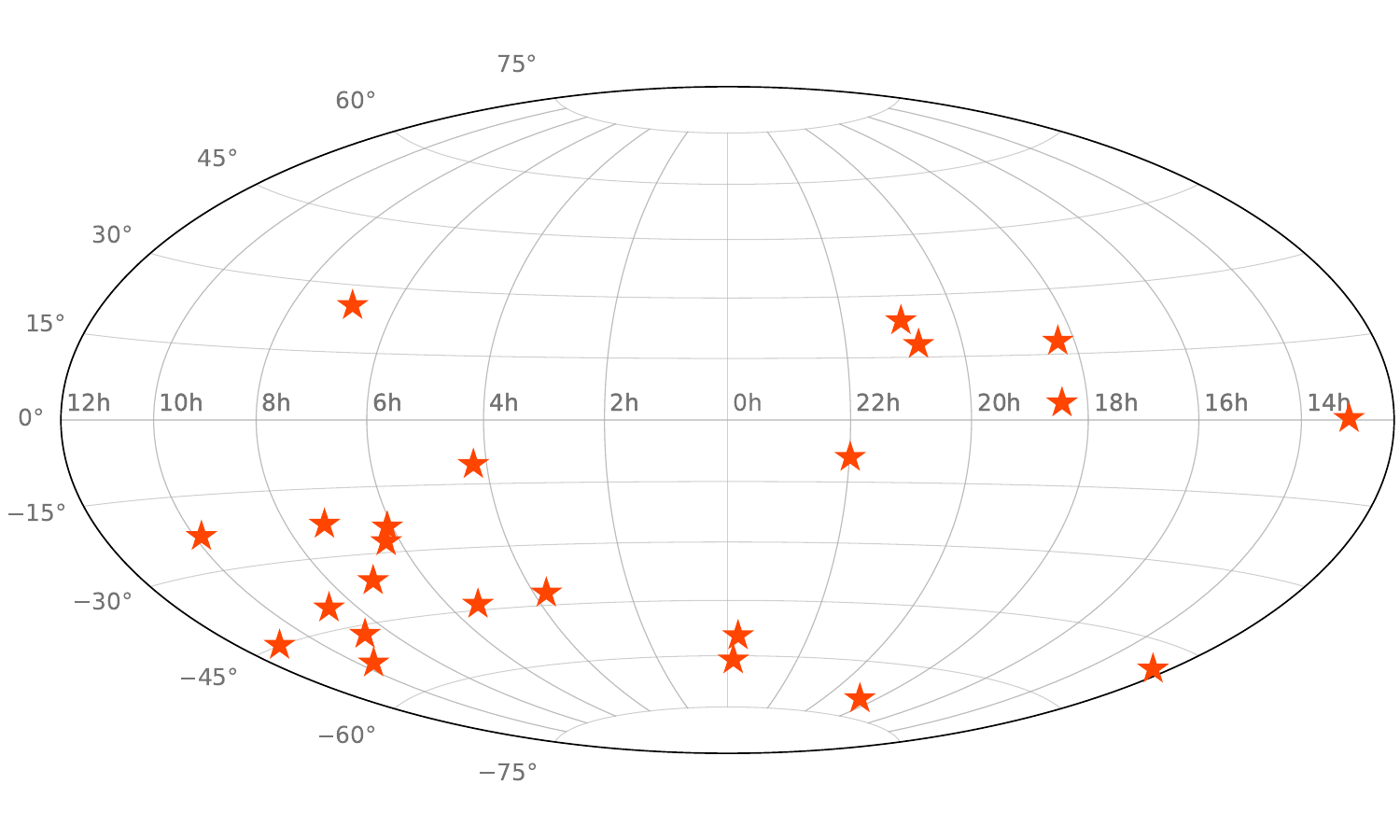}

\FigCap{On-sky view on bright LSP stars presented in equatorial coordinates in the Hammer projection.}
\label{fig:map}
\end{figure}

\begin{figure}[h]
\begin{center}
\includegraphics[scale=0.13]{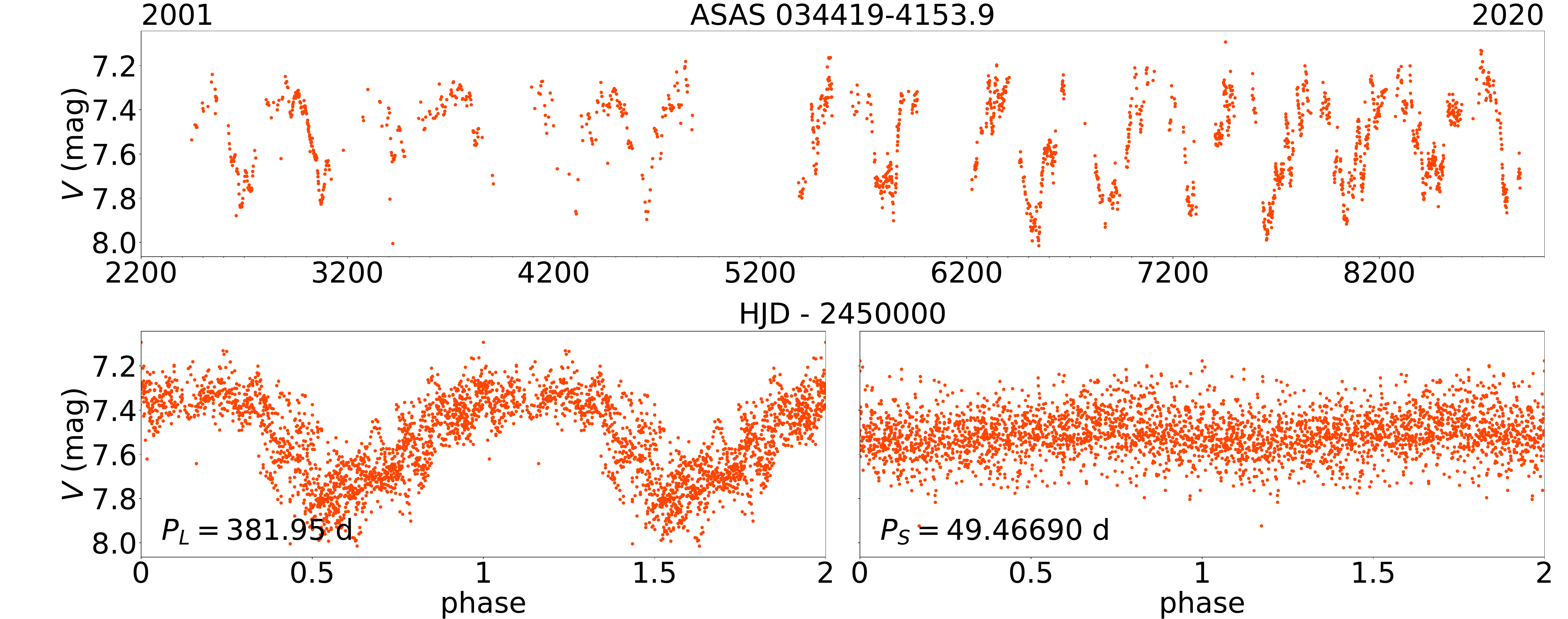}
\includegraphics[scale=0.13]{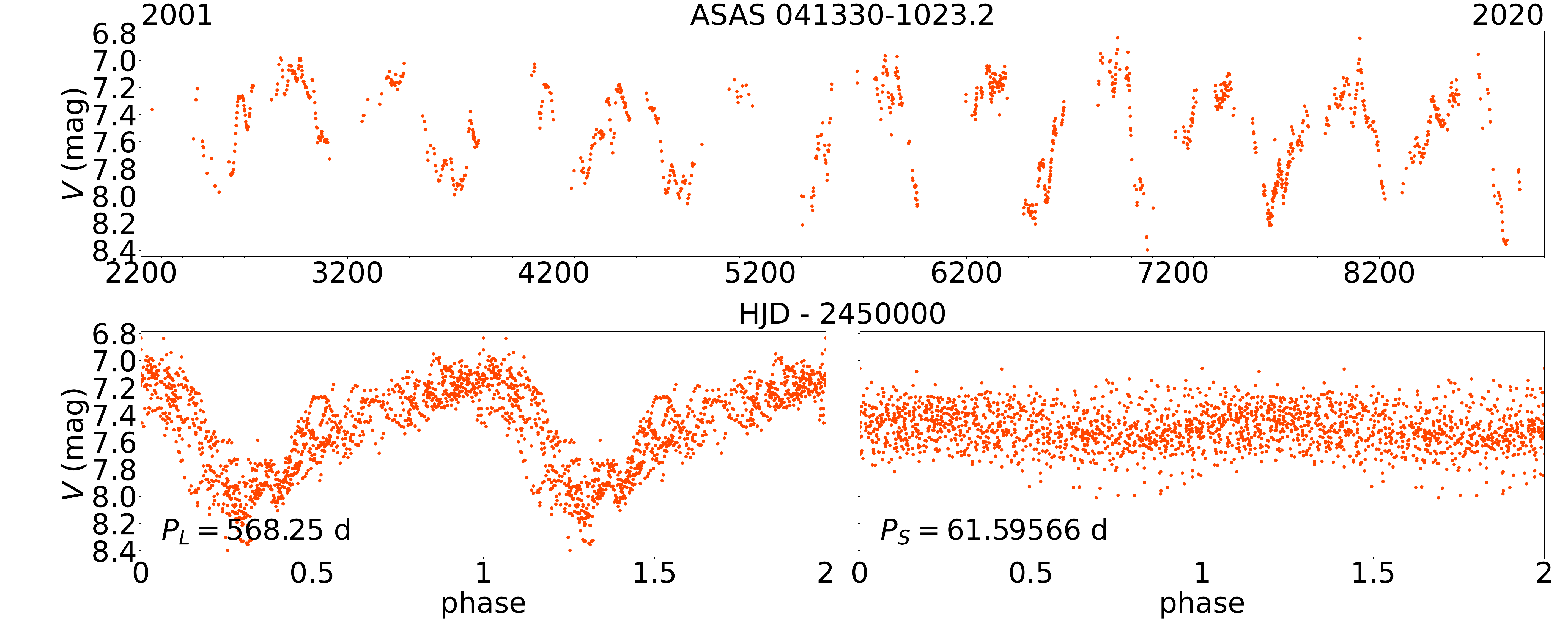}
\includegraphics[scale=0.13]{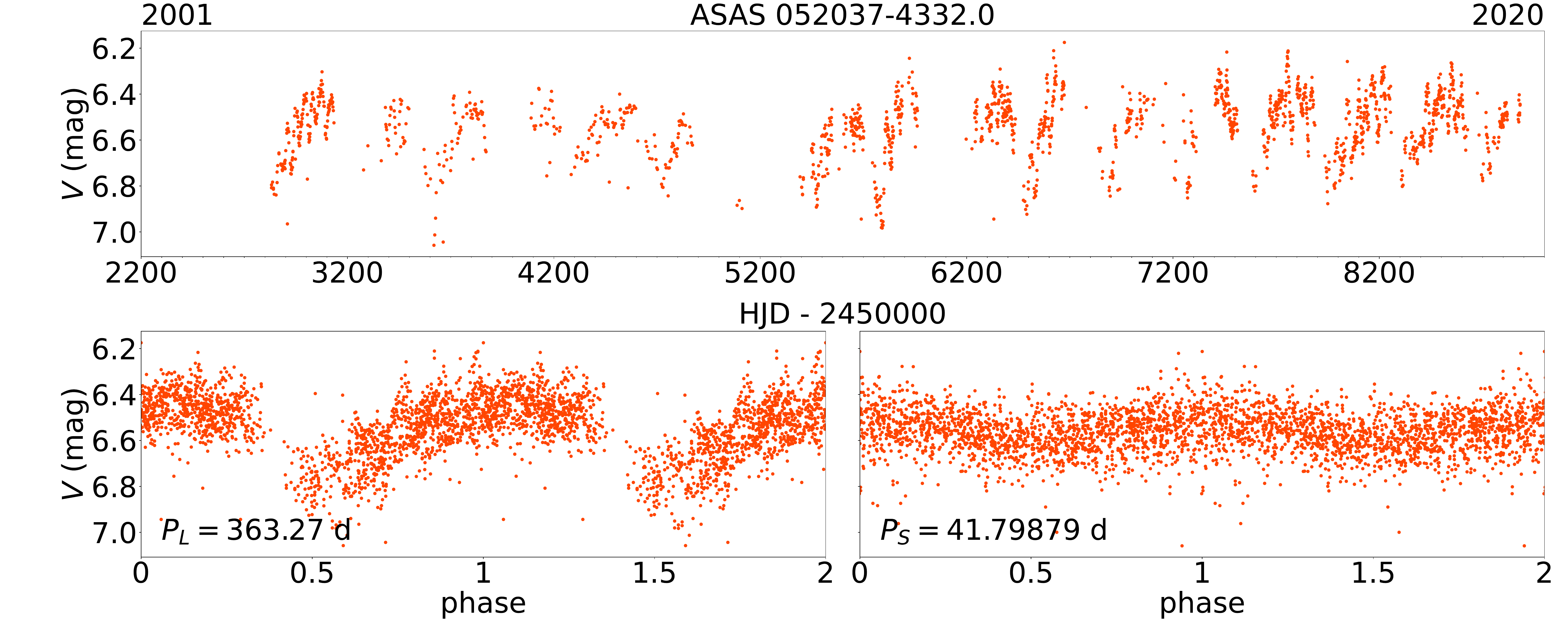}
\includegraphics[scale=0.13]{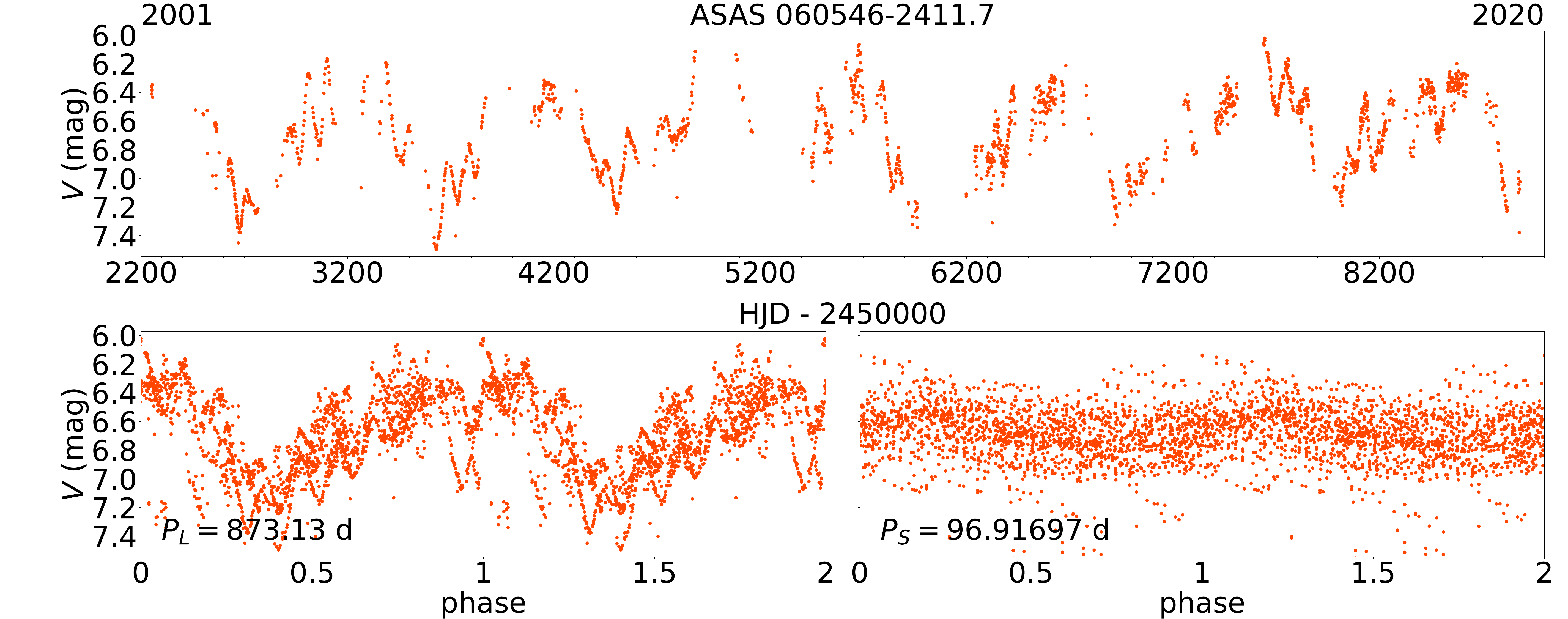}
\end{center}

\FigCap{Four examples of LSP stars discovered in the ASAS data. Each object is shown in three different panels. \textit{Top panel}: the full unfolded ASAS \textit{V}-band light curve with the first and last year of the observations marked above the panel together with the star's ID. \textit{Bottom left panel}: phase-folded light curve with the LSP $P_L$, indicated inside the plot. \textit{Bottom right panel}: phase-folded light curve with the pulsation period $P_S$, also indicated inside the plot.}
\label{fig:lc_part1}

\end{figure}

\begin{figure}[h]
\begin{center}
\includegraphics[scale=0.13]{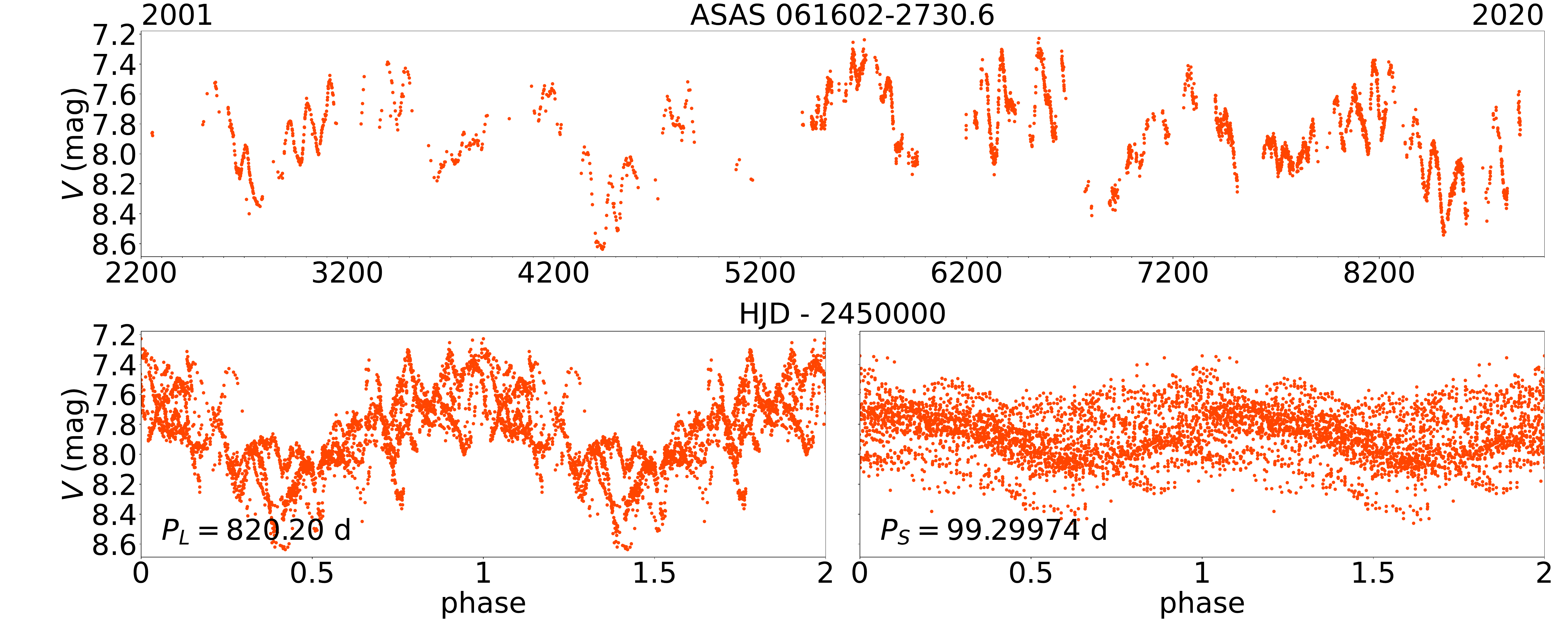}
\includegraphics[scale=0.13]{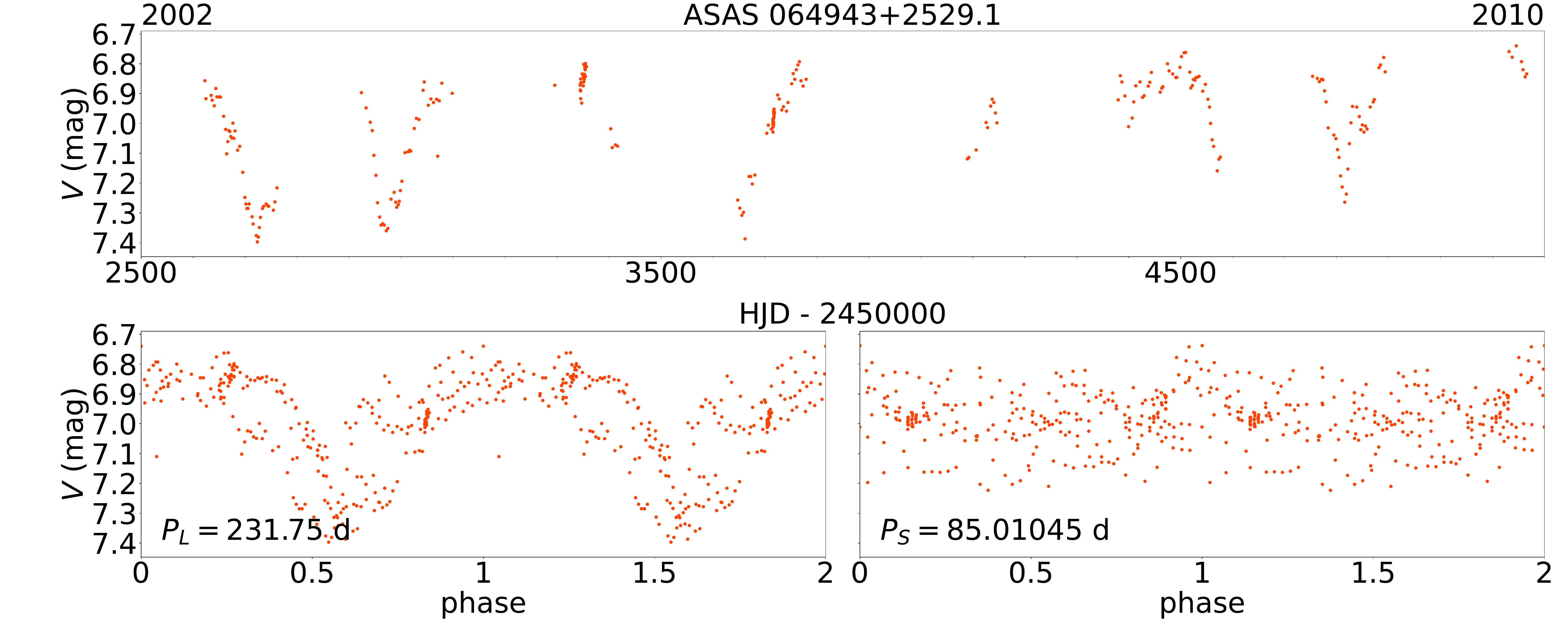}
\includegraphics[scale=0.13]{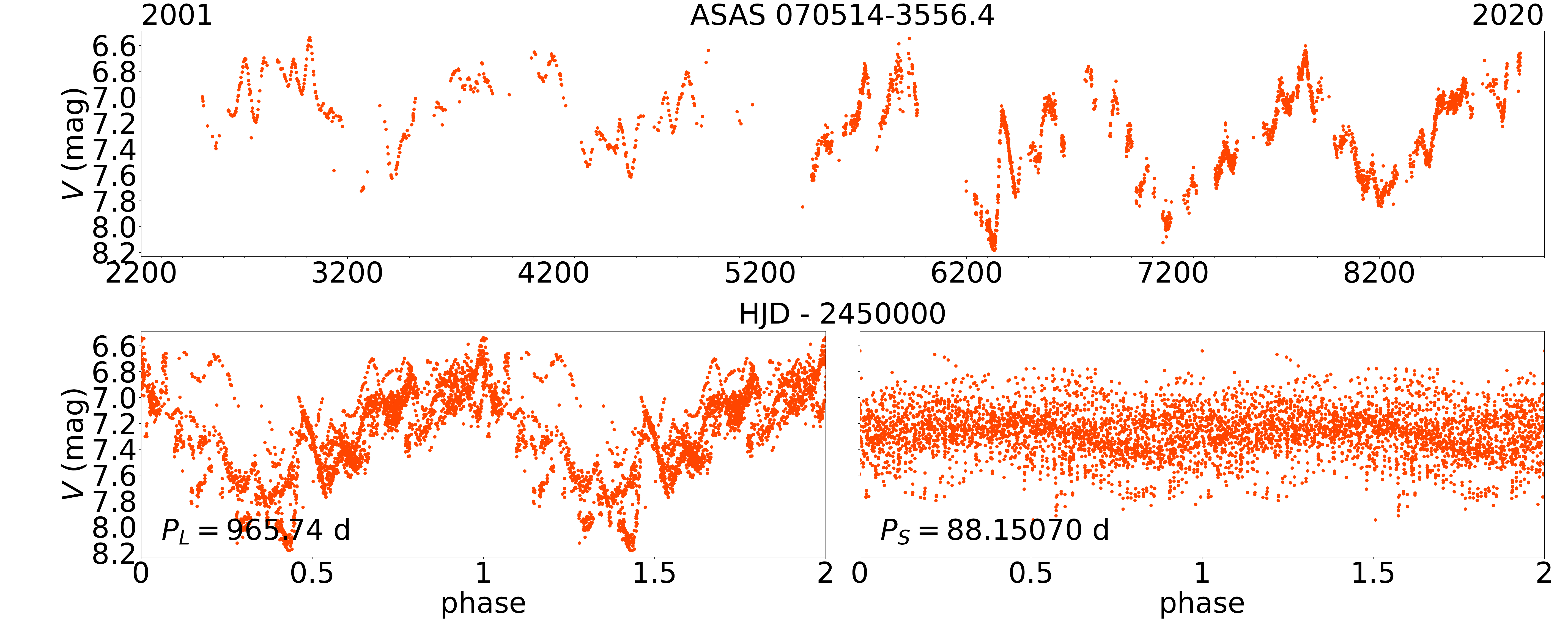}
\includegraphics[scale=0.13]{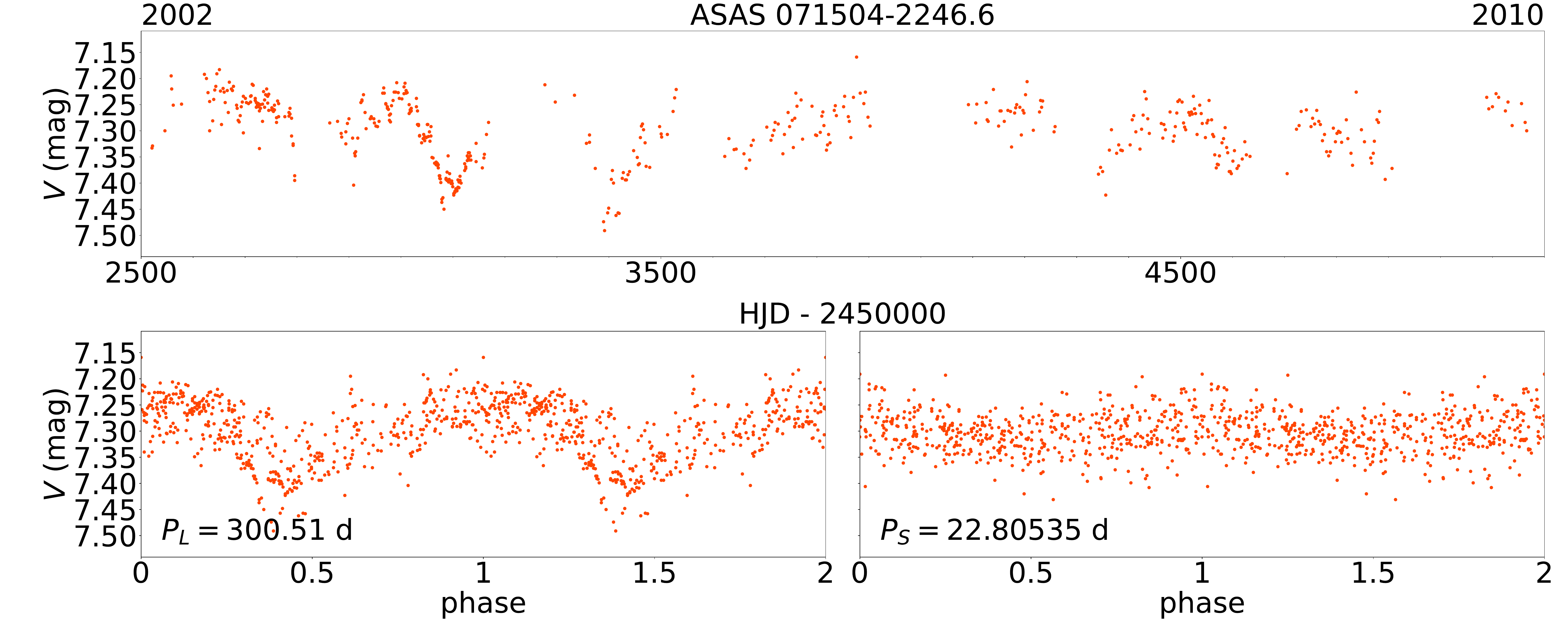}
\end{center}

\FigCap{Same as Fig. \ref{fig:lc_part1}, but four other examples are presented.}
\label{fig:lc_part2}

\end{figure}

\begin{figure}[h]
\begin{center}
\includegraphics[scale=0.13]{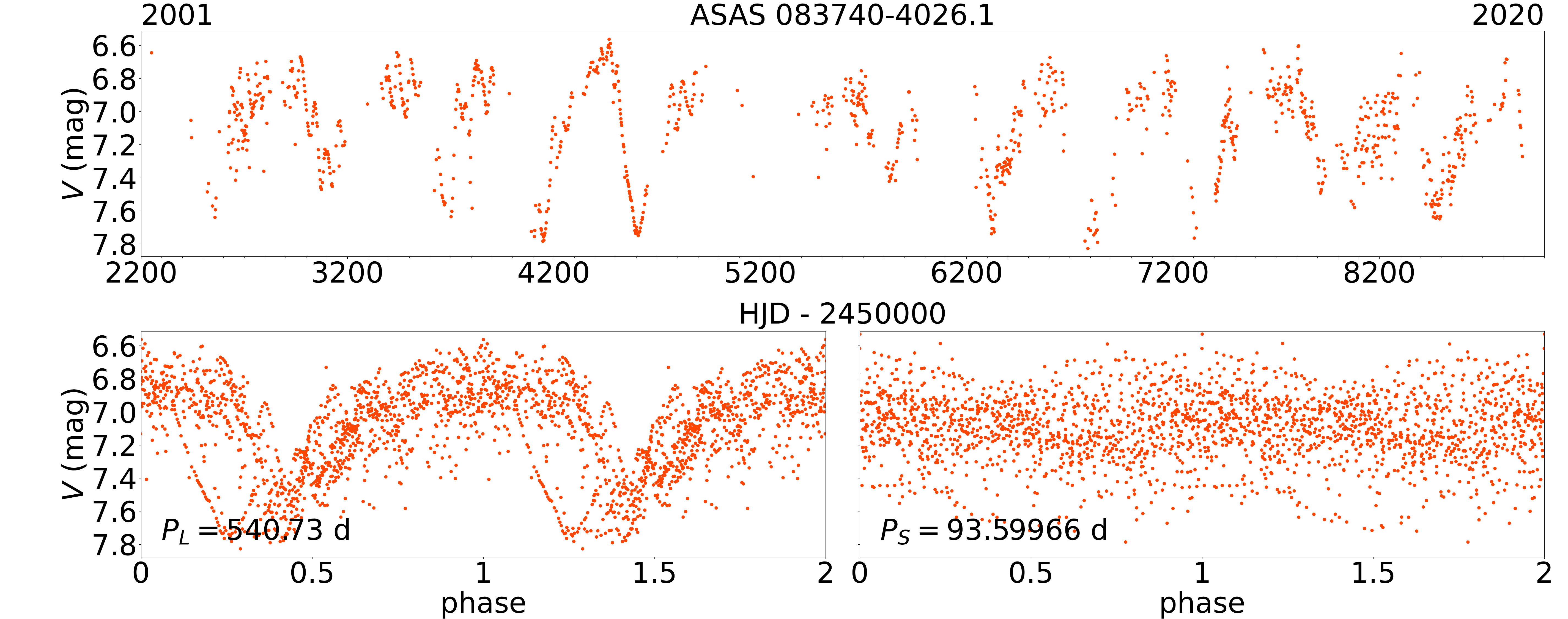}
\includegraphics[scale=0.13]{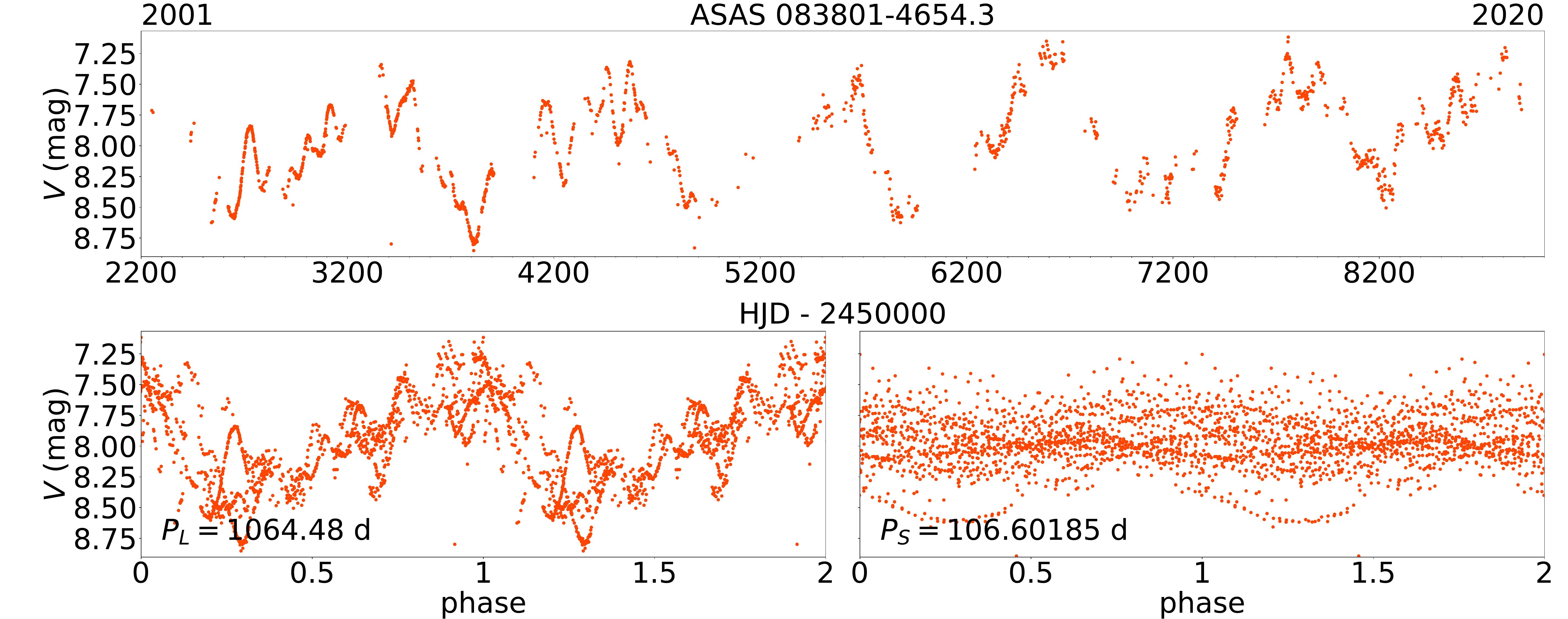}
\includegraphics[scale=0.13]{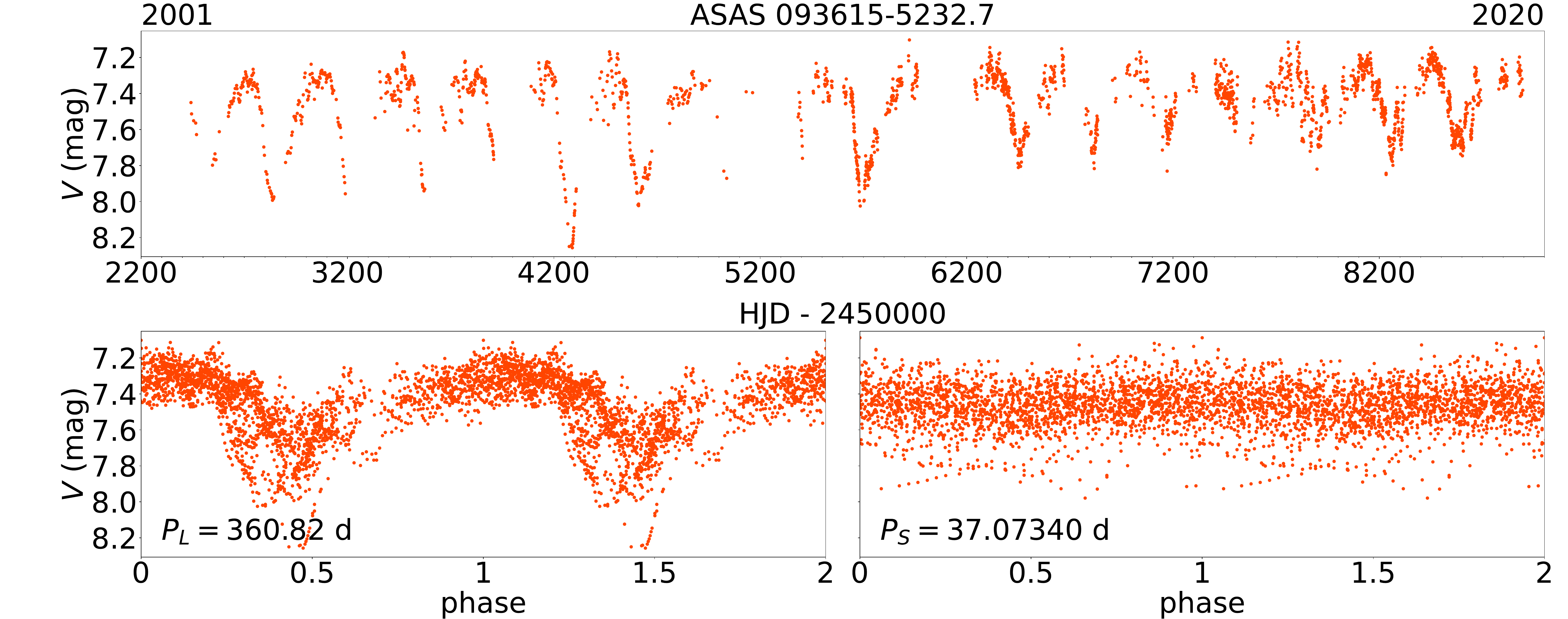}
\includegraphics[scale=0.13]{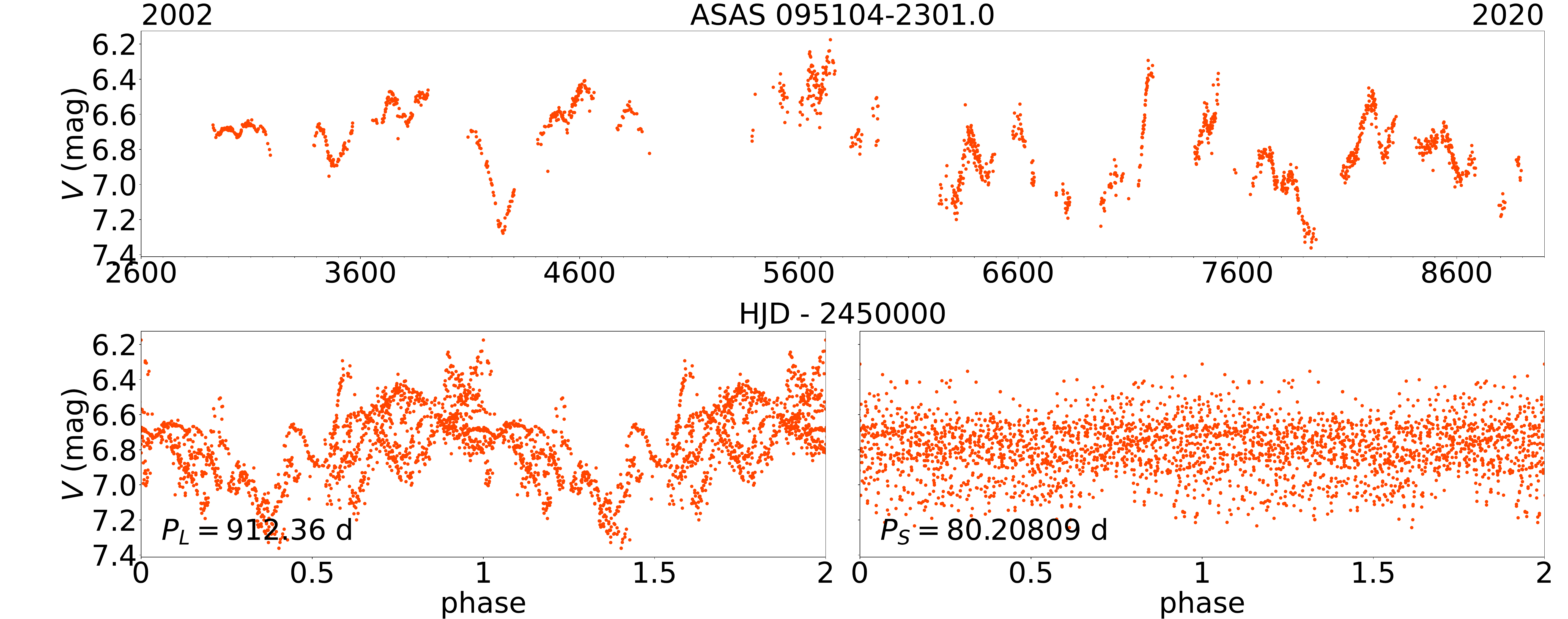}
\end{center}

\FigCap{Same as Fig. \ref{fig:lc_part1}, but four other examples are presented.}
\label{fig:lc_part3}

\end{figure}

\begin{figure}[h]
\begin{center}
\includegraphics[scale=0.13]{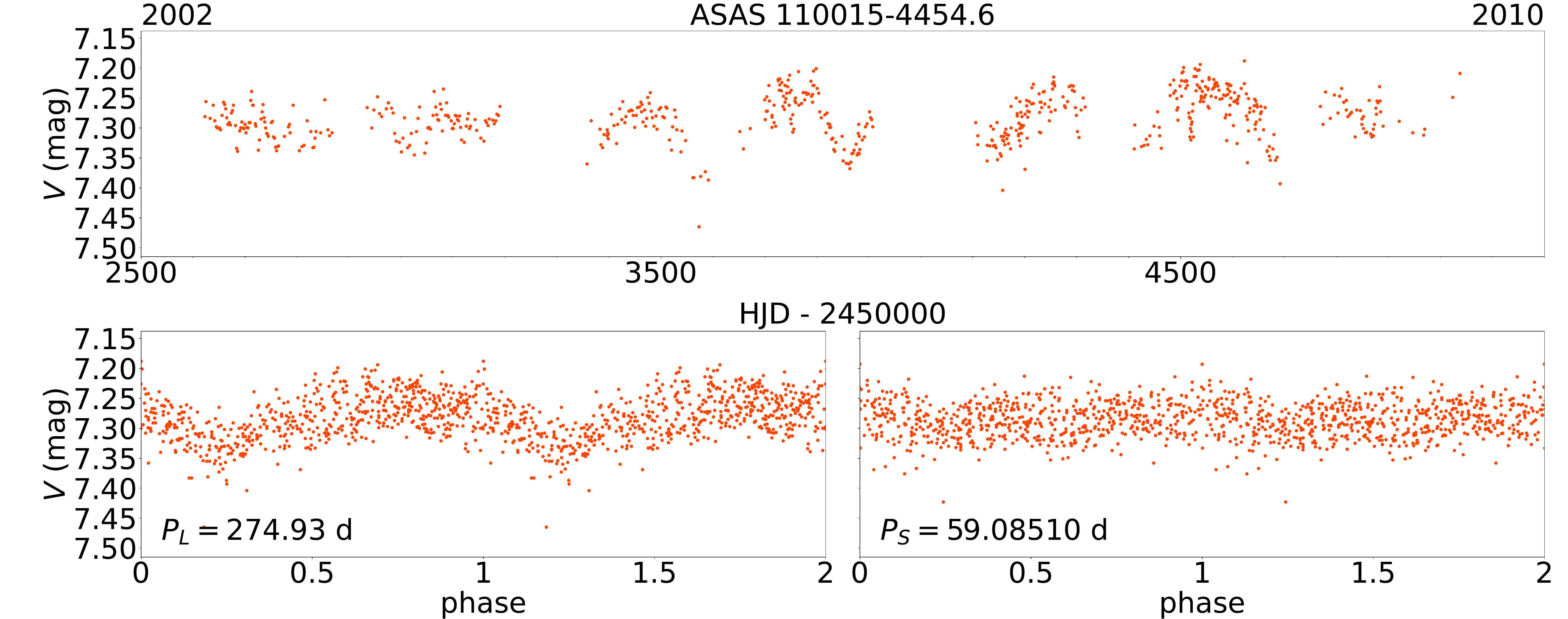}
\includegraphics[scale=0.13]{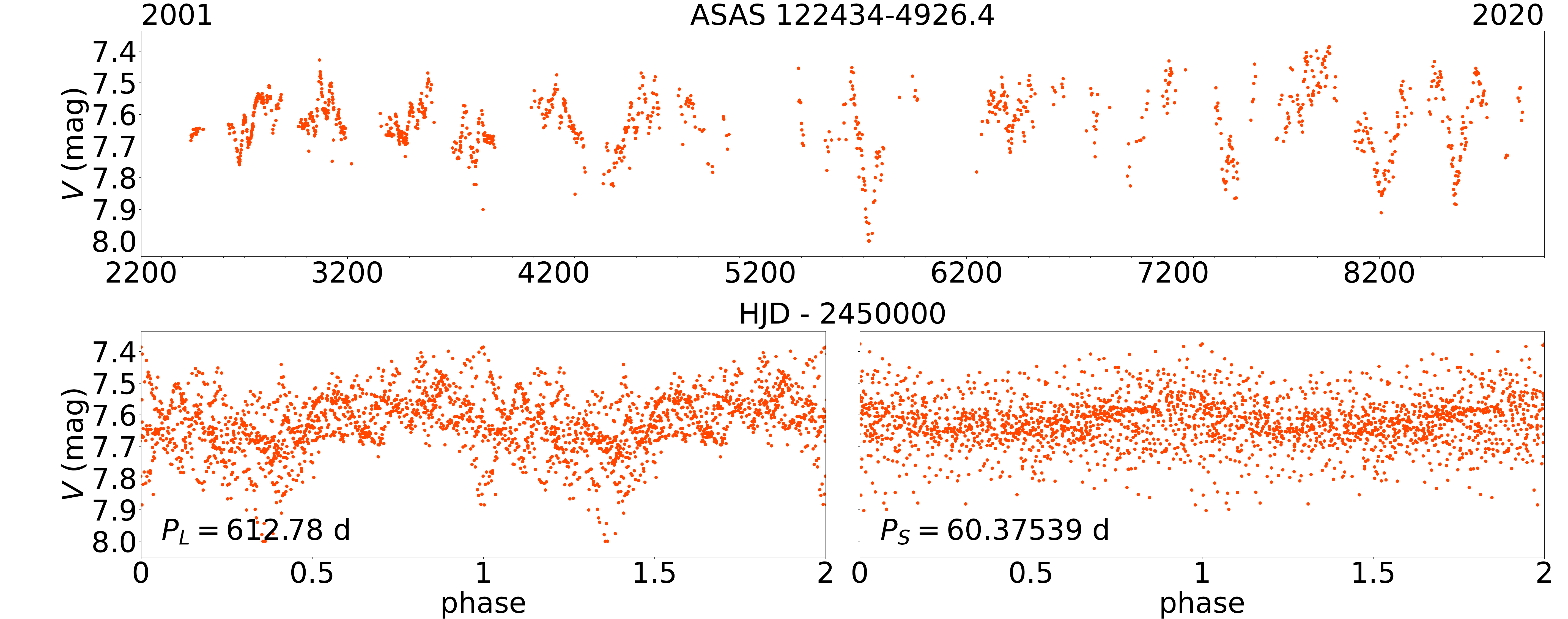}
\includegraphics[scale=0.13]{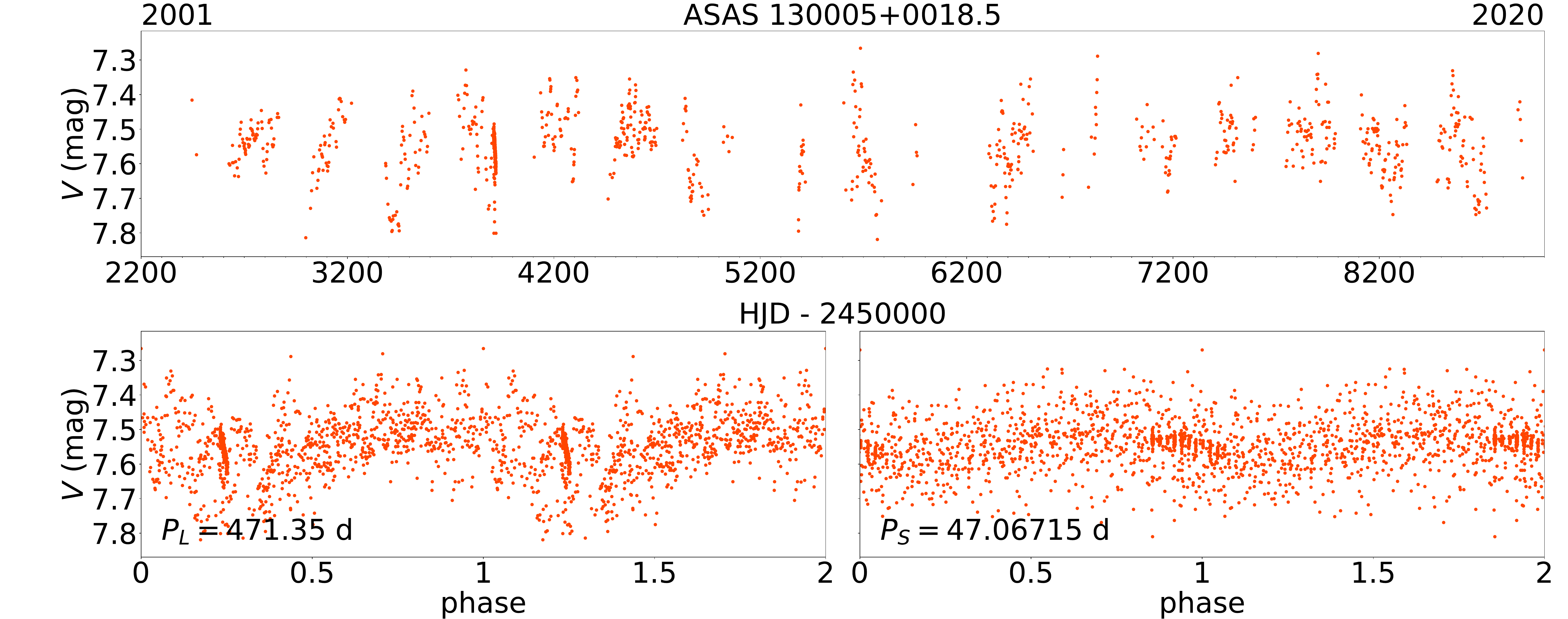}
\includegraphics[scale=0.13]{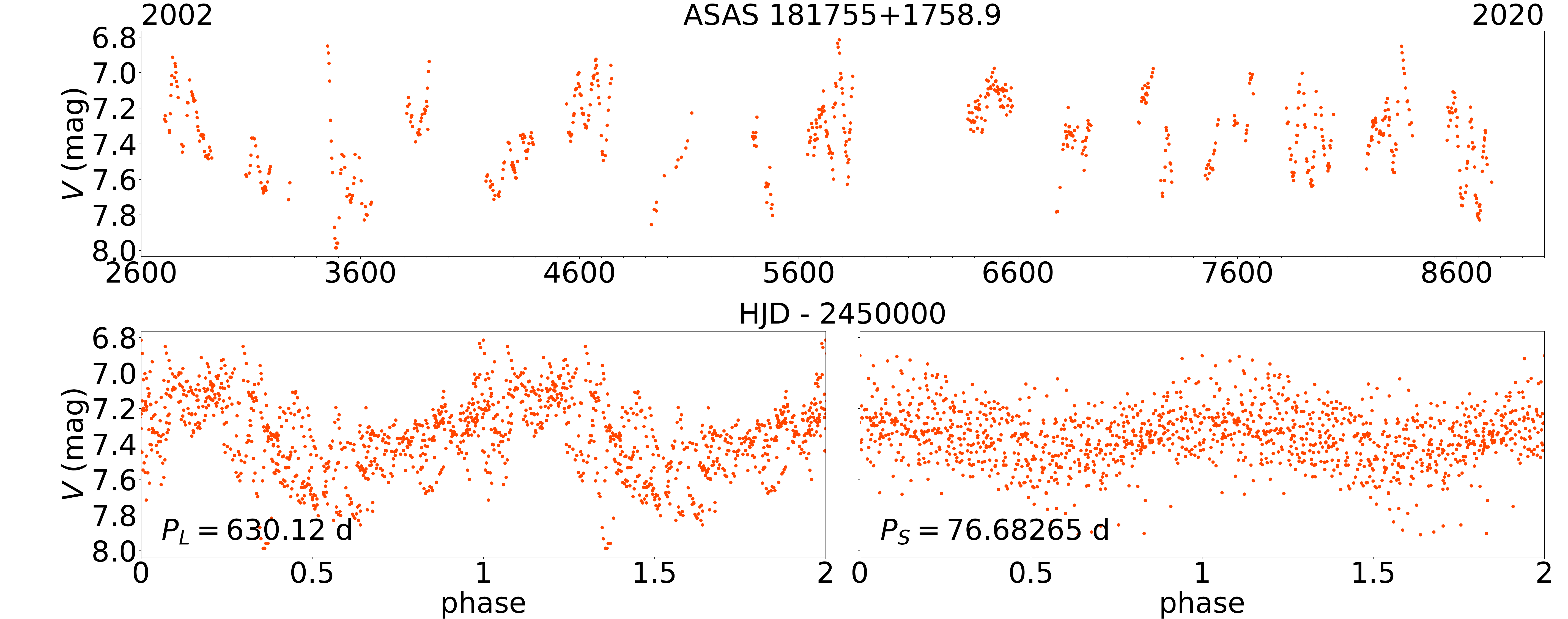}
\end{center}

\FigCap{Same as Fig. \ref{fig:lc_part1}, but four other examples are presented.}
\label{fig:lc_part4}

\end{figure}

\begin{figure}[h]
\begin{center}
\includegraphics[scale=0.13]{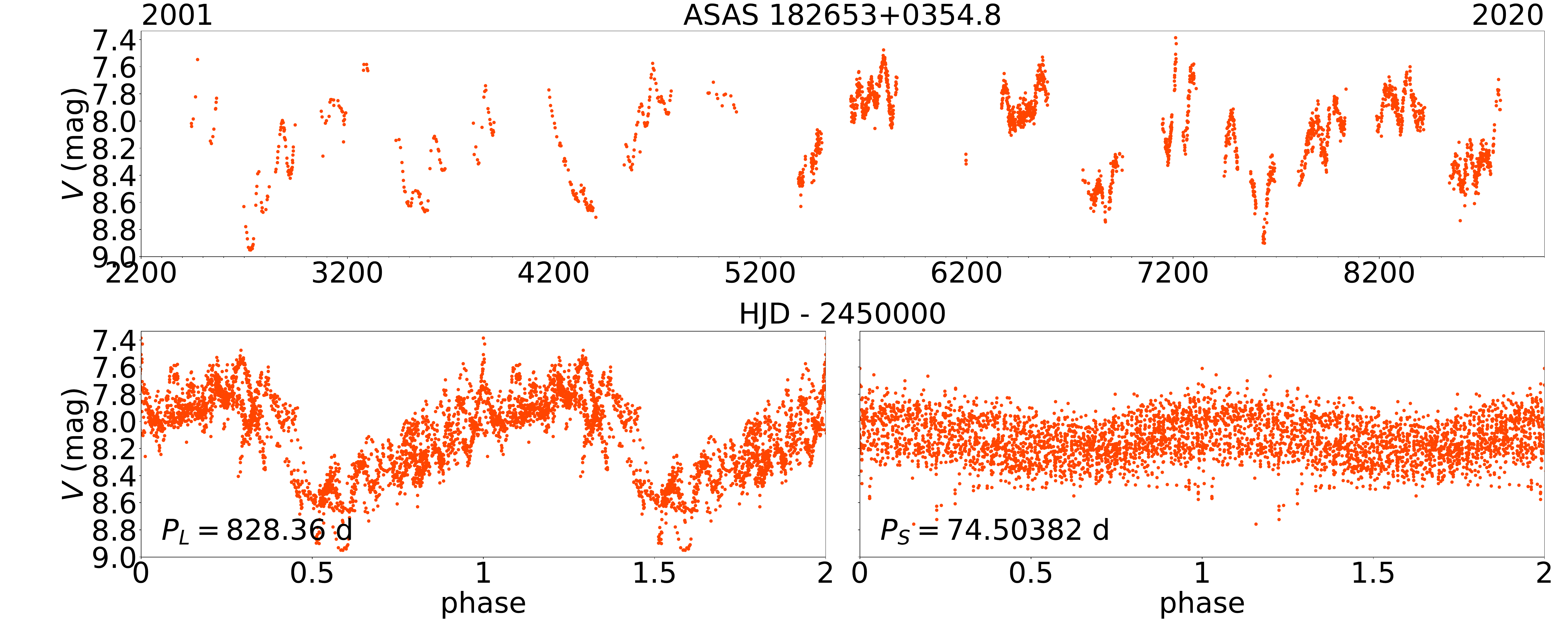}
\includegraphics[scale=0.13]{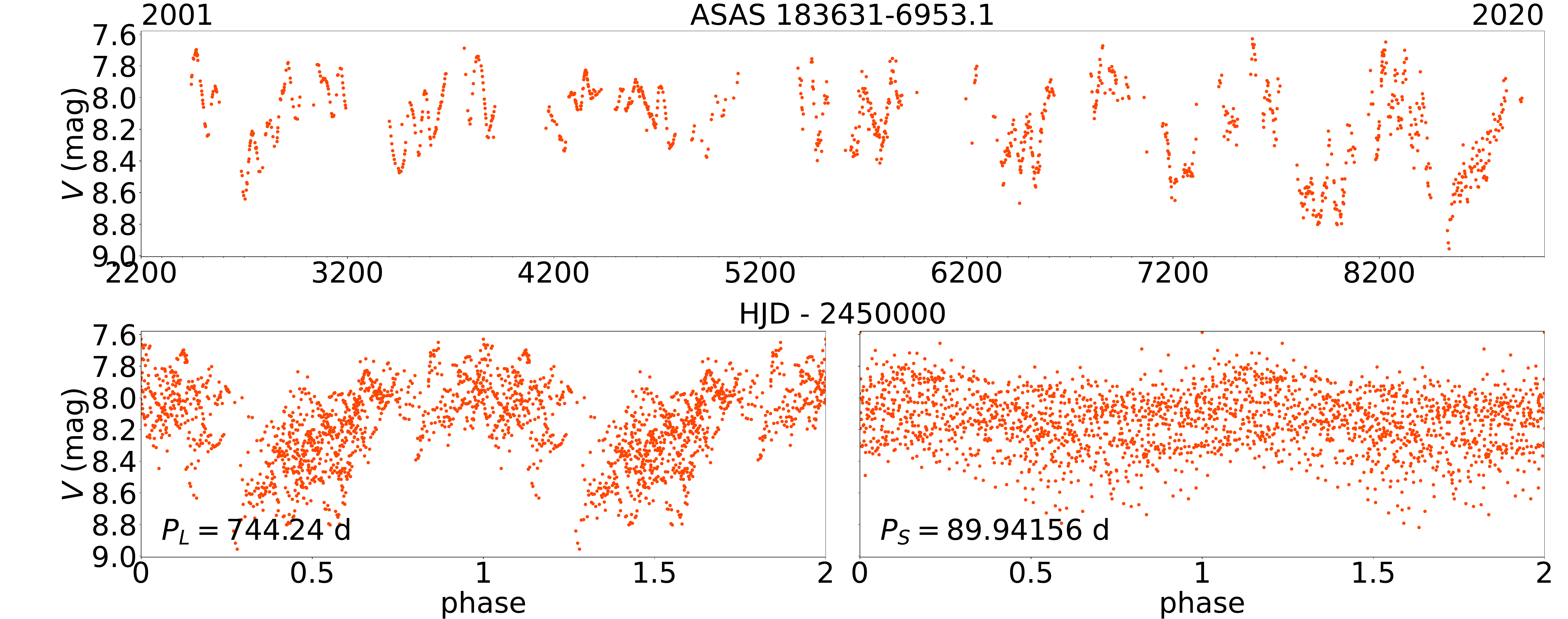}
\includegraphics[scale=0.13]{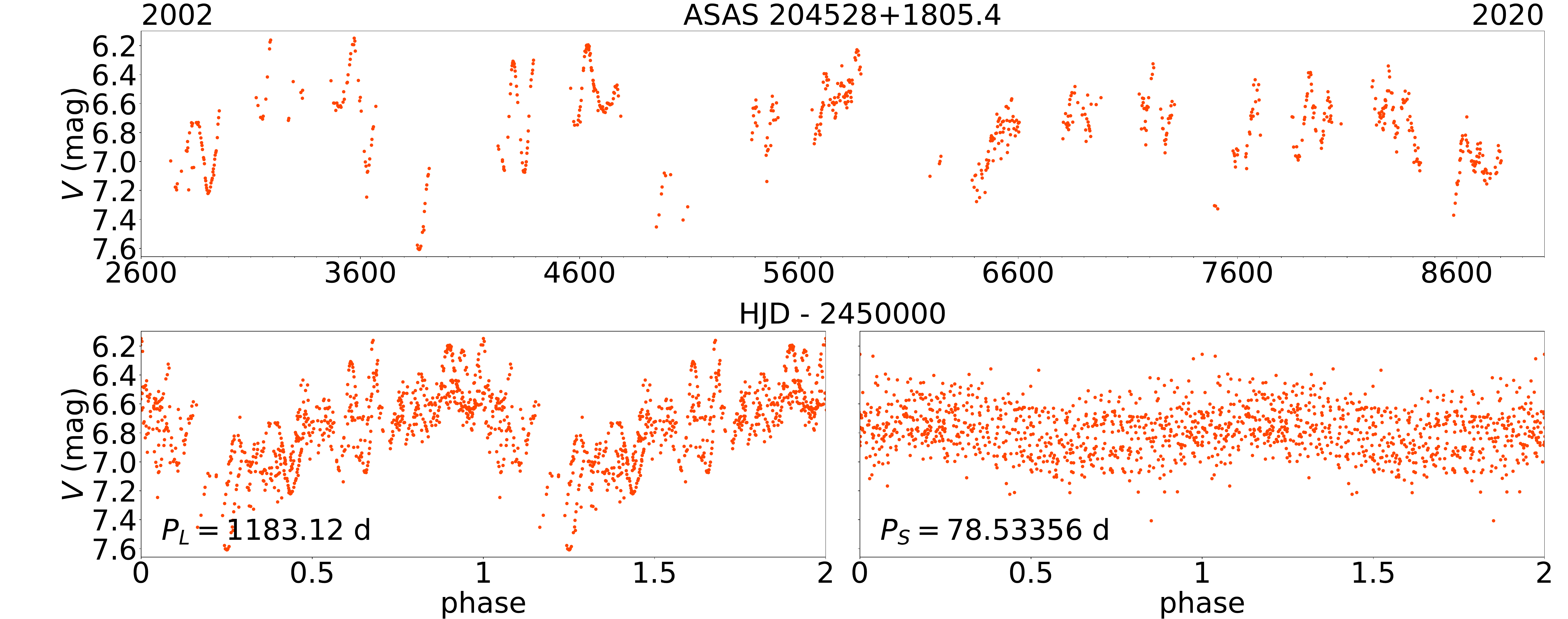}
\includegraphics[scale=0.13]{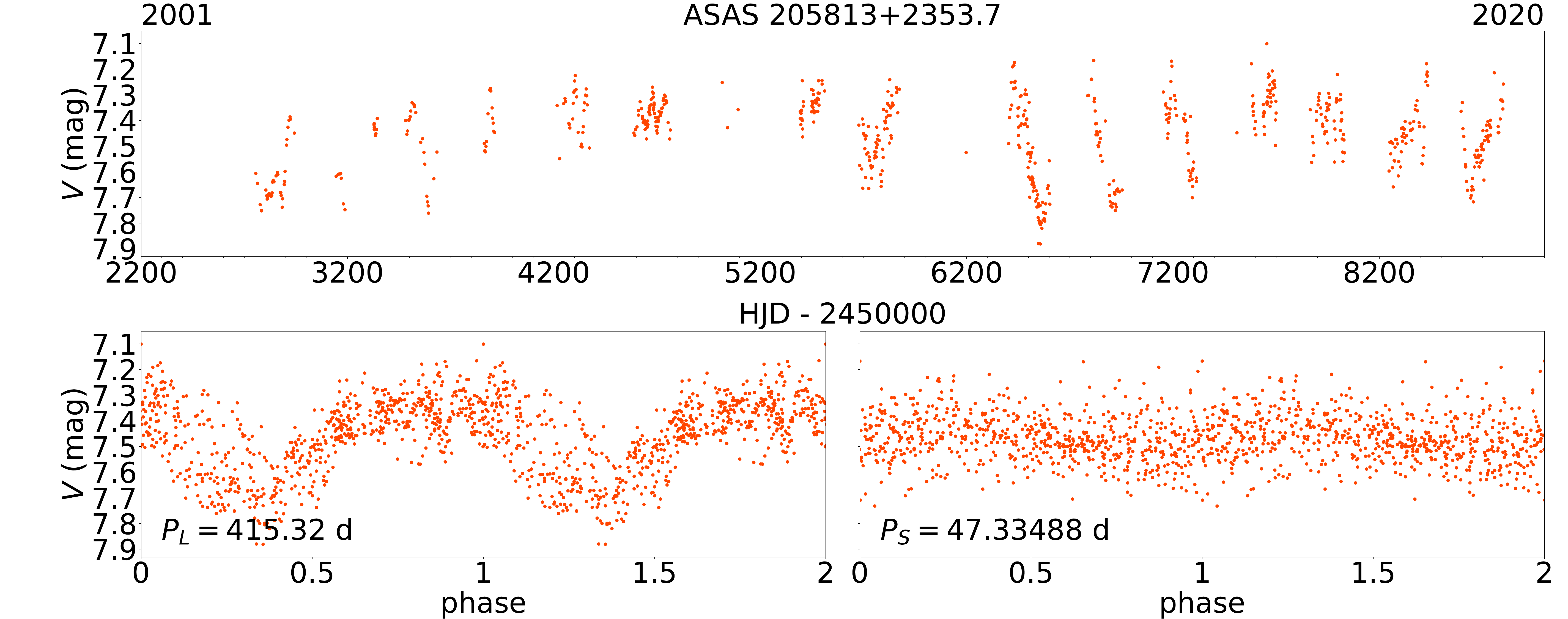}
\end{center}

\FigCap{Same as Fig. \ref{fig:lc_part1}, but four other examples are presented.}
\label{fig:lc_part5}

\end{figure}

\begin{figure}[h]
\begin{center}
\includegraphics[scale=0.13]{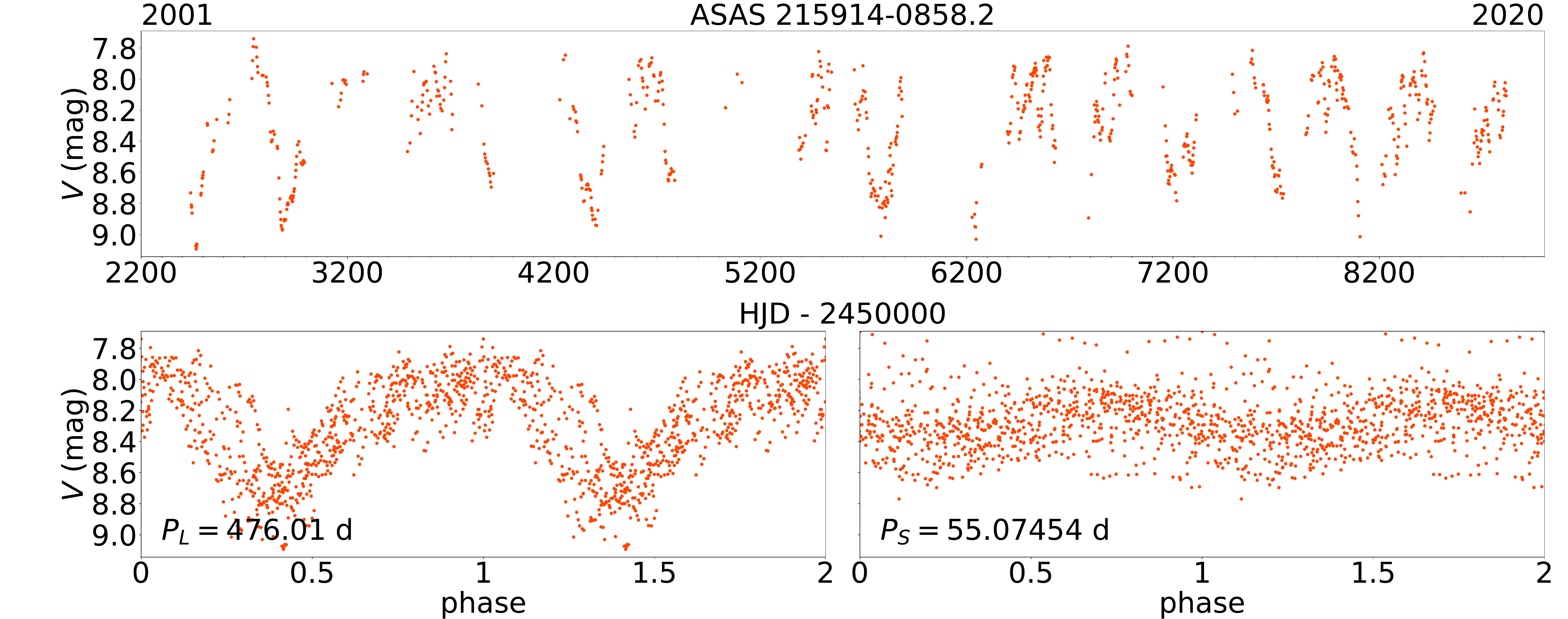}
\includegraphics[scale=0.13]{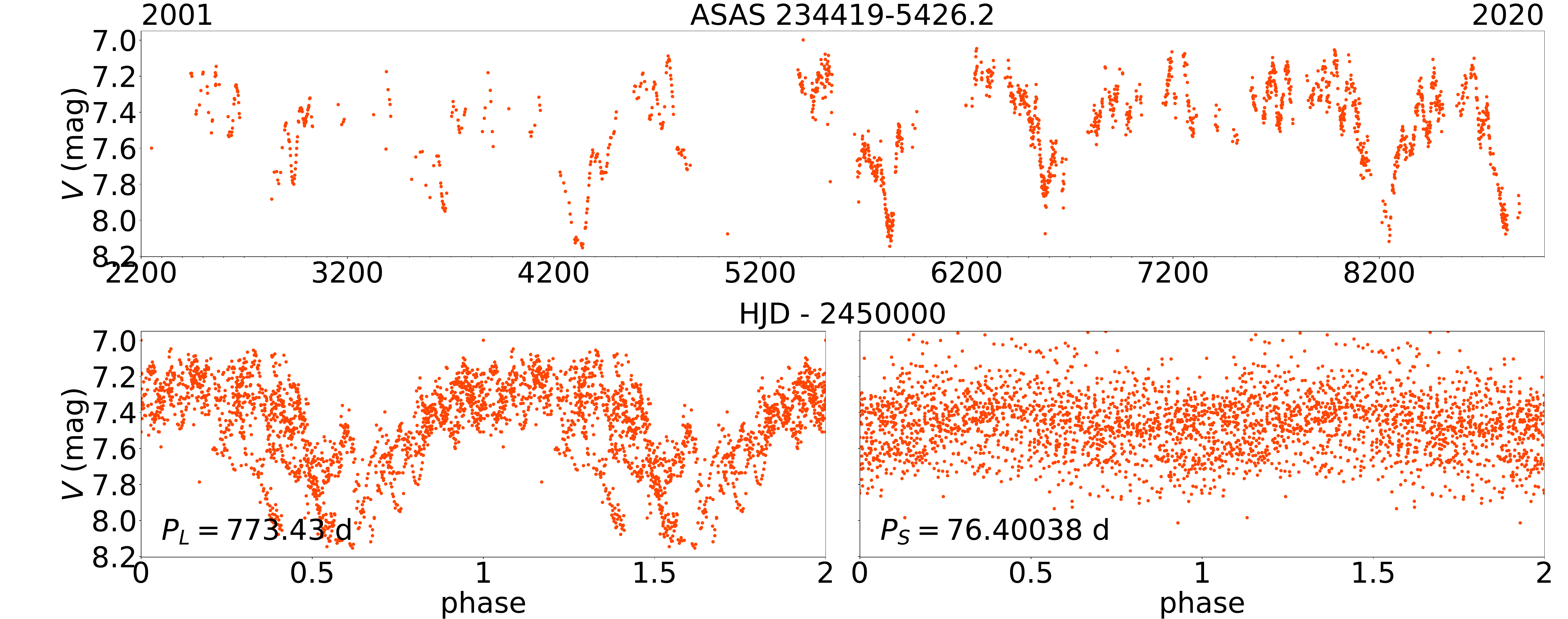}
\includegraphics[scale=0.13]{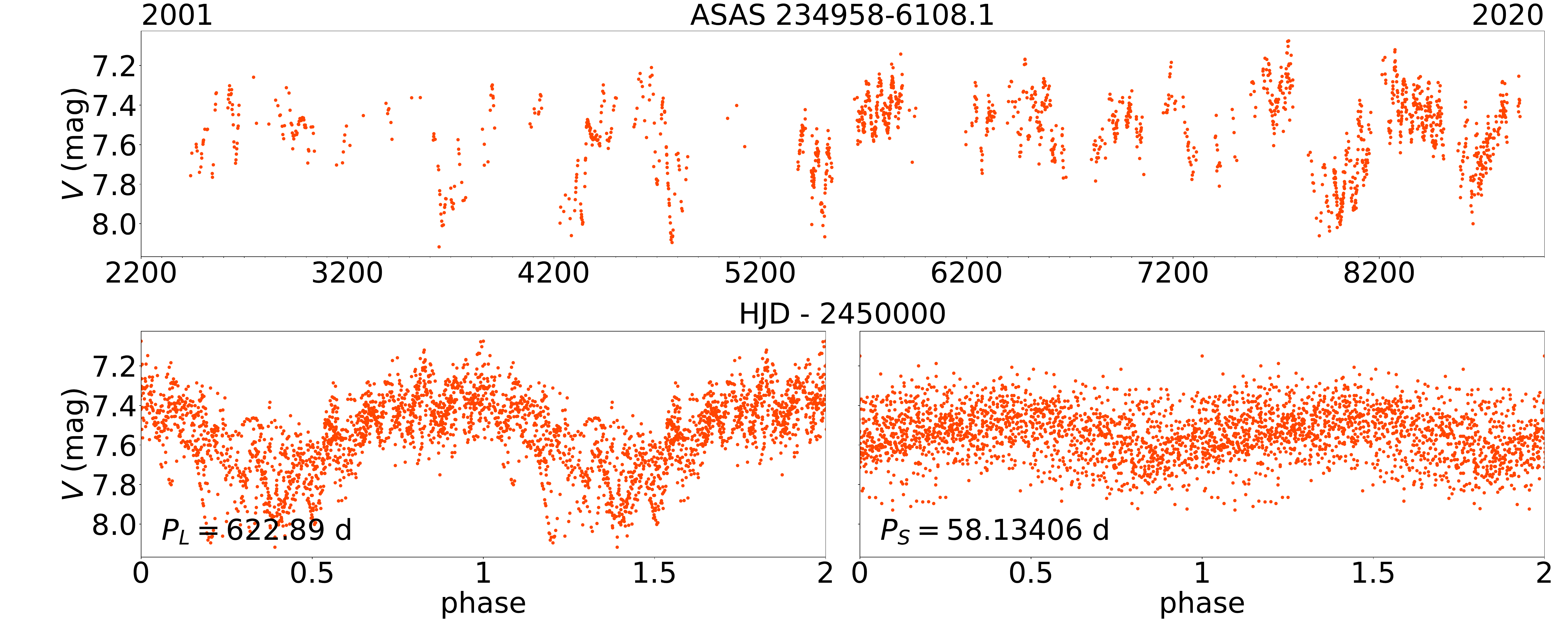}
\end{center}

\FigCap{Same as Fig. \ref{fig:lc_part1}, but three other examples are presented.}
\label{fig:lc_part6}

\end{figure}

The sky distribution of the sample is shown in Figure \ref{fig:map}, while the ASAS light curves (both unfolded and phase-folded with periods $P_L$ and $P_S$) are presented in Figures \ref{fig:lc_part1}, \ref{fig:lc_part2}, \ref{fig:lc_part3}, \ref{fig:lc_part4}, \ref{fig:lc_part5}, and \ref{fig:lc_part6} ordered by increasing right ascension. All ASAS \textit{V}-~band data used in this study, together with Table \ref{tab:lsp} in a machine-readable format, are available at:

\begin{itemize}
\item \url{https://www.astrouw.edu.pl/asas/?page=LSP} $\rightarrow$ summary of the paper,
\item \url{https://www.astrouw.edu.pl/asas/i_LSP/bright_lsp_asas.txt} $\rightarrow$ Table \ref{tab:lsp} in a machine-readable format,
\item \url{https://www.astrouw.edu.pl/asas/i_LSP/LSP.zip} $\rightarrow$ compressed file containing ASAS \textit{V}-band light curves.
\end{itemize}

The light curves are provided as individual files named according to the ASAS ID with the \texttt{.dat} extension (e.g. \texttt{034419-4153.9.dat}). Each light curve file contains three columns: HJD-2450000, \textit{V}-band magnitude, and the corresponding \textit{V}-band magnitude uncertainty.

We make this collection and its associated data publicly available to the astronomical community, as it provides an excellent sample for follow-up interferometric and spectroscopic observations aimed at probing the nature of the LSP phenomenon and advancing efforts to resolve this long-standing astrophysical puzzle.

\subsection{Crossmatch with other Catalogs}

We crossmatched our sample with the \textit{Gaia Data Release 3} \citep[Gaia DR3;][]{gaiadr32023}, the \textit{Two Micron All Sky Survey} (2MASS) Point Source Catalog \citep{skrutskie2006}, the \textit{International Variable Star Index} \citep[VSX;][]{watson2006}\footnote{\url{https://vsx.aavso.org/index.php} (accessed 16 December 2025)}, and the \textit{SIMBAD Astronomical Database} \citep{wenger2000}\footnote{\url{https://simbad.cds.unistra.fr/simbad/} (accessed 16 December 2025)} using a matching radius of $10$ arcsec. We found counterparts for all objects in each of the mentioned catalogs. Additionally, we crossmatched our

\begin{landscape}
\begin{table}[p]

\begin{adjustwidth}{-2cm}{0pt}
\caption{Table \texttt{bright\_lsp\_asas.txt} with all information about bright variables exhibiting LSP behaviour, i.e., ASAS, Gaia DR3 and other IDs, equatorial coordinates (R.A. and Decl.), mean ASAS \textit{V}-band magnitude (\textit{V}), \textit{V}-band amplitude ($A_V$), long secondary period ($P_L$), pulsation period ($P_S$), Gaia parallax ($\varpi$), parallax error ($\sigma_{\varpi}$), distance ($d$), distance error ($\sigma_d$), the Renormalized Unit Weight Error (RUWE), and \textit{JH$K_s$}-bands magnitudes from 2MASS. The table is ordered by increasing right ascension.}
\vspace{0.5cm}
\label{tab:lsp}
\setlength{\tabcolsep}{4pt}
\renewcommand{\arraystretch}{1.3}
\scriptsize

\begin{tabular}{lcccccccccccccccl}
\hline \hline
ASAS ID & R.A. & Decl. & $P_L$ & $P_S$ & \textit{V} & $A_V$ & Gaia DR3 ID & $\varpi$ & $\sigma_{\varpi}$ & $d$ & $\sigma_d$ & RUWE & \textit{J} & \textit{H} & \textit{$K_S$} & Other ID  \\
 & (h:m:s) & (deg:m:s) & (d) & (d) & (mag) & (mag) &  & (mas) & (mas) & (pc) & (pc) & & (mag) & (mag) & (mag) &  \\
\hline
034419-4153.9 & 03:44:19 & $-$41:53:54 & 381.95 & 49.46690 & 7.514 & 0.576 & 4848874011097672576 & 1.3433 & 0.0152 & 740 & 9 & 1.017 & 4.295 & 3.355 & 3.123 & EU Eri \\
041330-1023.2 & 04:13:30 & $-$10:23:12 & 568.25 & 61.59566 & 7.456 & 0.993 & 3191955033656822656 & 2.4857 & 0.0486 & 396 & 8 & 1.143 & 2.482 & 1.635 & 1.355 & BM Eri \\
052037-4332.0 & 05:20:37 & $-$43:32:00 & 363.27 & 41.79879 & 6.520 & 0.446 & 4800503505295394304 & 2.6781 & 0.0262 & 372 & 4 & 0.928 & 2.774 & 1.842 & 1.630 & WW Pic \\
060546-2411.7 & 06:05:46 & $-$24:11:42 & 873.13 & 96.91697 & 6.672 & 0.903 & 2913152694837288192 & 3.8975 & 0.1041 & 257 & 8 & 1.271 & 0.796 & $-$0.175 & $-$0.505 & S Lep \\
061602-2730.6 & 06:16:02 & $-$27:30:48 & 820.20 & 99.29974 & 7.860 & 0.869 & 2899323552978274176 & 2.3803 & 0.0523 & 415 & 9 & 1.222 & 2.033 & 1.087 & 0.787 & NSV 16816 \\
064943+2529.1 & 06:49:43 & $+$25:29:06 & 231.75 & 85.01045 & 6.960 & 0.511 & 3381600414124419584 & 1.8846 & 0.0367 & 521 & 10 & 1.135 & 4.094 & 3.344 & 3.099 & QU Gem \\
070514-3556.4 & 07:05:14 & $-$35:56:24 & 965.74 & 88.15070 & 7.215 & 1.151 & 5578428769590816256 & 3.6266 & 0.1542 & 277 & 12 & 3.752 & 0.674 & $-$0.240 & $-$0.544 & HD 53917 \\
071504-2246.6 & 07:15:04 & $-$22:46:36 & 300.51 & 22.80535 & 7.290 & 0.178 & 2927947792030206976 & 1.4405 & 0.1213 & 689 & 54 & 6.932 & 3.813 & 2.816 & 2.599 & HD 56159 \\
083740-4026.1 & 08:37:40 & $-$40:26:06 & 540.73 & 93.59966 & 7.043 & 0.909 & 5528303889906184960 & 2.3605 & 0.0269 & 419 & 4 & 0.972 & 2.530 & 1.621 & 1.346 & GO Vel \\
083801-4654.3 & 08:38:01 & $-$46:54:18 & 1064.48 & 106.60185 & 7.957 & 1.195 & 5521580651198875008 & 2.8405 & 0.1145 & 358 & 15 & 1.013 & 1.101 & 0.192 & $-$0.186 & MN Vel \\
093615-5232.7 & 09:36:15 & $-$52:32:42 & 360.82 & 37.07340 & 7.410 & 0.604 & 5312562975355731456 & 1.8384 & 0.0158 & 542 & 4 & 0.994 & 3.963 & 2.981 & 2.755 & MS Vel \\
095104-2301.0 & 09:51:04 & $-$23:01:00 & 912.36 & 80.20809 & 6.745 & 0.684 & 5663852920326093952 & 2.1991 & 0.0961 & 453 & 20 & 0.932 & 2.255 & 1.124 & 0.521 & Y Hya \\
110015-4454.6 & 11:00:15 & $-$44:54:36 & 274.93 & 59.08510 & 7.282 & 0.118 & 5387245515953555968 & 1.6454 & 0.0346 & 597 & 12 & 1.842 & 3.887 & 2.935 & 2.727 & V0358 Vel \\
122434-4926.4 & 12:24:34 & $-$49:26:24 & 612.78 & 60.37539 & 7.624 & 0.318 & 6126837819170763008 & 1.5635 & 0.0291 & 638 & 12 & 0.989 & 3.933 & 2.848 & 2.442 & S Cen \\
130005+0018.5 & 13:00:05 & $+$00:18:30 & 471.35 & 47.06715 & 7.539 & 0.273 & 3689410721541718272 & 1.5102 & 0.0306 & 646 & 14 & 1.017 & 4.083 & 3.103 & 2.815 & HD 112915 \\
181755+1758.9 & 18:17:55 & $+$17:58:54 & 630.12 & 76.68265 & 7.338 & 0.701 & 4523659121810239744 & 4.2058 & 0.0744 & 237 & 4 & 1.811 & 1.285 & 0.248 & 0.070 & IQ Her \\
182653+0354.8 & 18:26:53 & $+$03:54:48 & 828.36 & 74.50382 & 8.031 & 0.910 & 4284072754993455616 & 2.2172 & 0.1170 & 436 & 24 & 1.031 & 2.103 & NULL & 0.827 & V0988 Oph \\
183631-6953.1 & 18:36:31 & $-$69:53:06 & 744.24 & 89.94157 & 8.133 & 0.824 & 6431818980098135040 & 1.9790 & 0.0657 & 506 & 16 & 0.825 & 2.685 & 1.719 & 1.388 & RT Pav \\
204528+1805.4 & 20:45:28 & $+$18:05:24 & 1183.12 & 78.53356 & 6.716 & 0.862 & 1813013108479261312 & 2.9865 & 0.1006 & 328 & 12 & 1.060 & 1.026 & $-$0.019 & $-$0.212 & U Del \\
205813+2353.7 & 20:58:13 & $+$23:53:42 & 415.32 & 47.33488 & 7.425 & 0.463 & 1839669874543519872 & 1.3914 & 0.0221 & 705 & 10 & 0.997 & 4.111 & 3.201 & 2.888 & HD 199696 \\
215914-0858.2 & 21:59:14 & $-$08:58:12 & 476.01 & 55.07454 & 8.260 & 0.930 & 2617670489740842496 & 1.4052 & 0.0344 & 689 & 15 & 1.079 & 3.244 & 2.255 & 1.966 & HD 208843\\
234419-5426.2 & 23:44:19 & $-$54:26:12 & 773.43 & 76.40038 & 7.425 & 0.813 & 6497754390309381376 & 1.9502 & 0.1017 & 515 & 28 & 1.698 & 3.244 & 2.255 & 1.966 & CI Phe \\
234958-6108.1 & 23:49:58 & $-$61:08:06 & 622.89 & 58.13406 & 7.515 & 0.630 & 6487905510607409280 & 2.2205 & 0.0238 & 446 & 5 & 0.989 & 2.583 & 1.706 & 1.440 & DU Tuc \\

\hline \hline
\end{tabular}

\begin{tablenotes}
\footnotesize
\item Note: The table is sorted by increasing right ascension and is available in its entirety in a machine-readable form through the website: \newline \url{https://www.astrouw.edu.pl/asas/i_LSP/bright_lsp_asas.txt}.
\end{tablenotes}

\end{adjustwidth}

\end{table}
\end{landscape}

\noindent sample with the distance catalog of \citet{bailer-jones2021}, from which we retrieved the photogeometric distances $d$. The uncertainties of the distances, $\sigma_d$, were calculated as the mean of the absolute differences between the retrieved distances and the 16th and 84th percentiles of the posterior distance distribution. We did not crossmatch our sample with mid-IR sky surveys such as \textit{Wide-field Infrared Survey Explorer} (WISE) or \textit{Spitzer}, as these objects are too bright to have reliable mid-IR measurements.

Therefore, Table \ref{tab:lsp} also lists Gaia DR3 and other catalog identifiers, the parallax ($\varpi$) and its uncertainty ($\sigma_{\varpi}$), the distance ($d$) and distance uncertainty ($\sigma_d$), the Renormalized Unit Weight Error (RUWE), as well as \textit{JH$K_S$}-band magnitudes. In cases where RUWE $> 1.4$, the parallax measurement may be unreliable and should be interpreted with caution.

\subsection{Comparison with \citet{percy2023}} \label{sec:percy}

In 2023, \citeauthor{percy2023} published a study of bright red giants and supergiants based on data from the American Association of Variable Star Observers (AAVSO) Binocular Observing Program. In that work, $111$ stars were analyzed for periodic variability, including possible LSP variability. The authors identified the LSP in $63$ objects in their sample, while $48$ classified as non-LSP stars.

We found that $47$ out of $111$ objects from their Tables 1, 2, and 3 are present in the ASAS database (after applying our selection criteria on \textit{V}-band magnitude, color index $(B-V)$, and declination $< 28^\circ$). Among them, $20$ stars are listed by \citet{percy2023} as not showing any LSP behavior and $27$ are classified as LSPs. We examined all these stars in detail using their ASAS light curves.  In the case of $20$ non-LSP stars we confirm that they indeed show no evidence of LSP variability. These stars are: RT~Cnc, RT~Cap, S~Car, BO~Car, R~Cen, V744~Cen, T~Cet, $\omicron$~Cet, R~Dor, RR~Eri, $\pi$~1~Gru, R~Hya, U~Hya, T~Mic, Z~Psc, L2~Pup, AH~Sco, Y~Tau, X~TrA, DM~Tuc.

In contrast, we found that only $8$ out of $27$ objects, that were classified as LSP variables by \citeauthor{percy2023}, show the presence of LSP variability in the ASAS data, while the remaining $19$ stars show a different type of long-term variability. Based on the ASAS light curves, we confirm the LSP in BM~Eri, S~Lep, GO~Vel, MN~Vel, Y~Hya, IQ~Her, U~Del, CI~Phe. These eight objects are already included in our collection. The 19 stars for which \citet{percy2023} found LSP, but we do not see it the ASAS data, are: $\theta$~Aps, W~CMa, CT~Del, EU~Del, Z~Eri, BR~Eri, TV~Gem, BU~Gem, T~Ind, RX~Lep, RV~Mon, BO~Mus, W~Ori, BL~Ori, Y~Pav, GO~Peg, TV~Psc, BM~Sco, FP~Vir.

\begin{figure}[h]

\begin{center}
\includegraphics[scale=0.13]{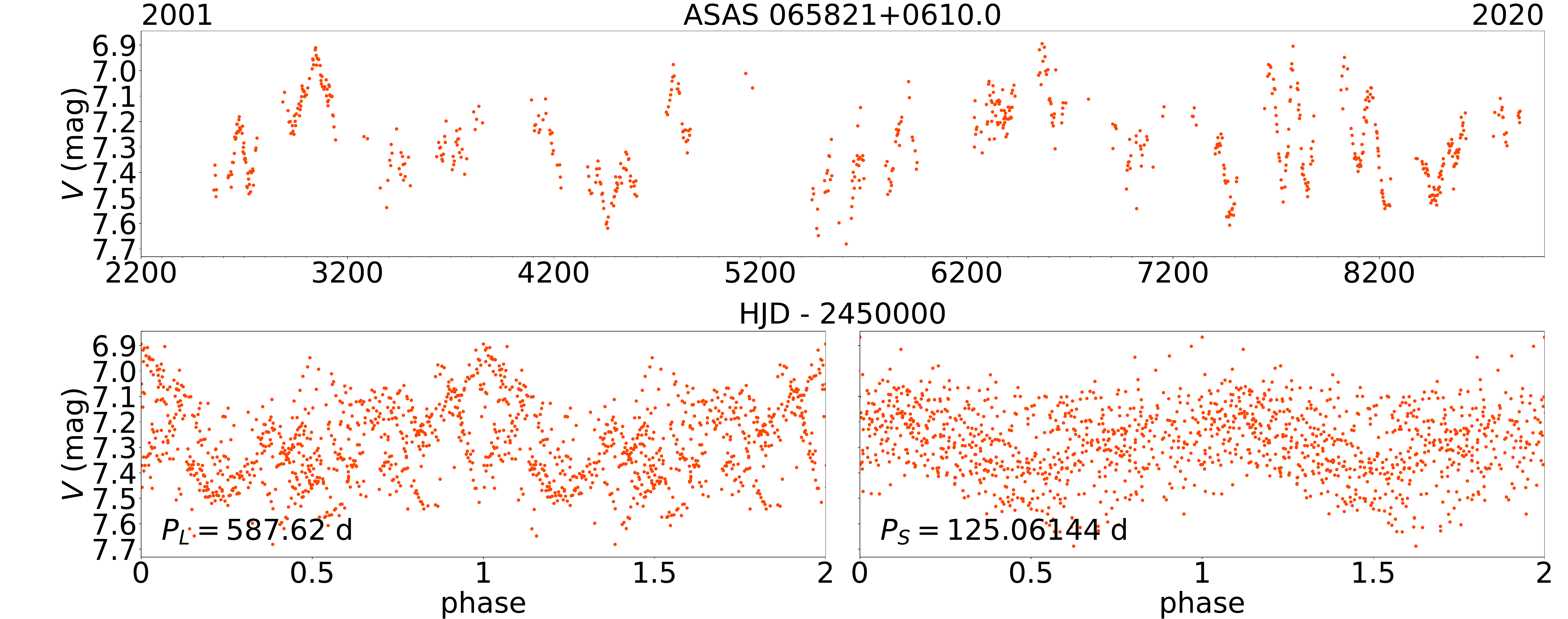}
\includegraphics[scale=0.13]{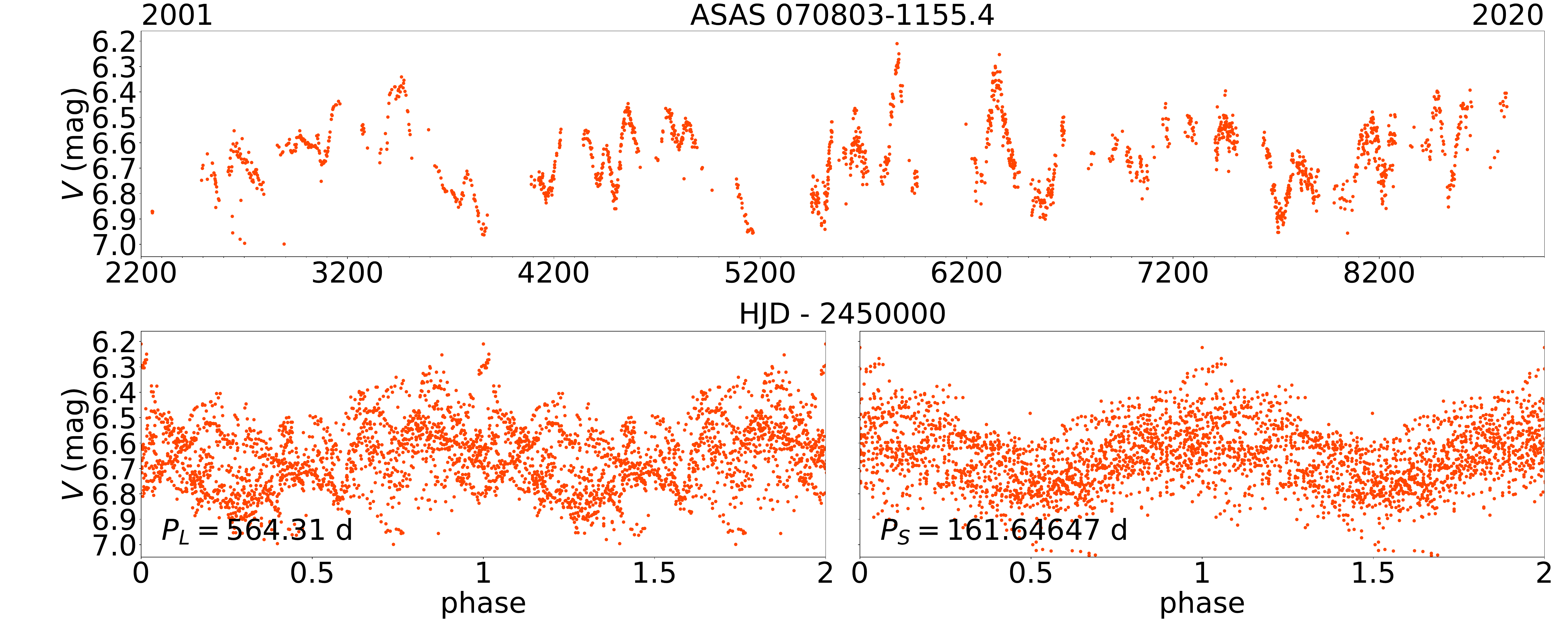}
\includegraphics[scale=0.13]{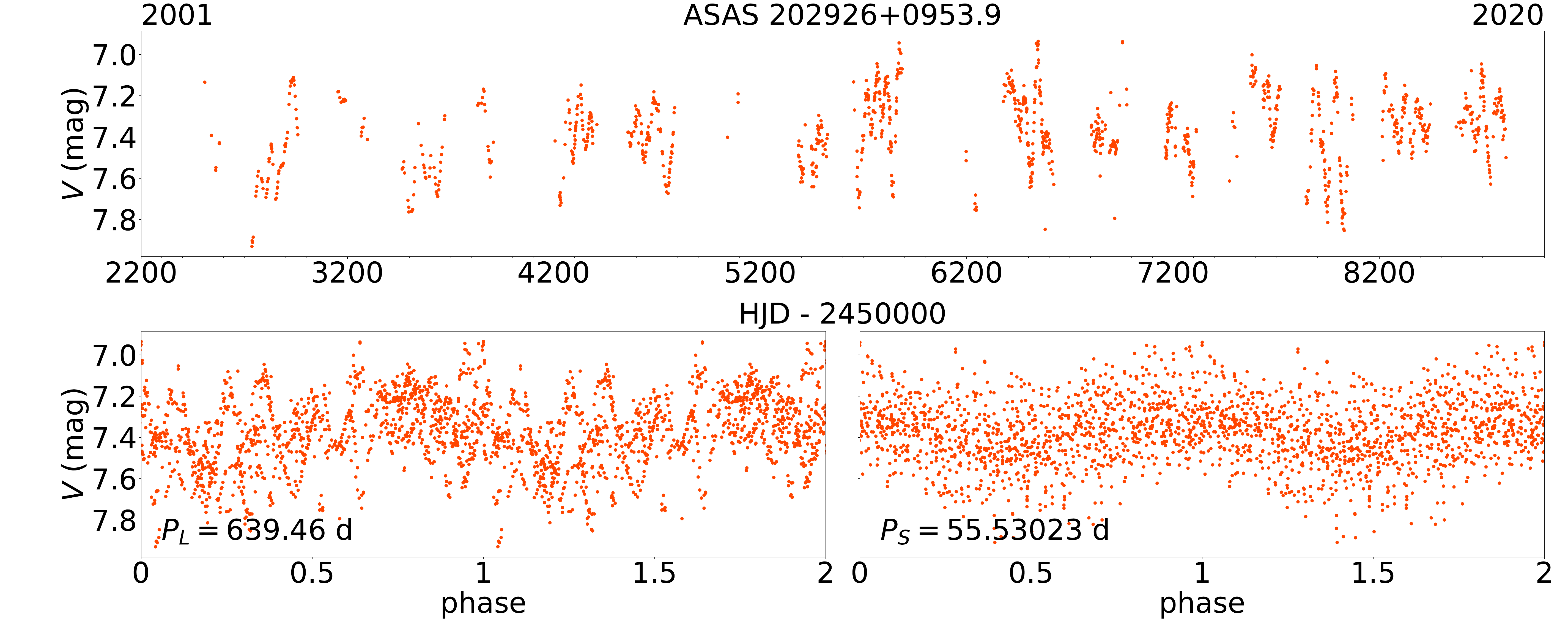}
\includegraphics[scale=0.13]{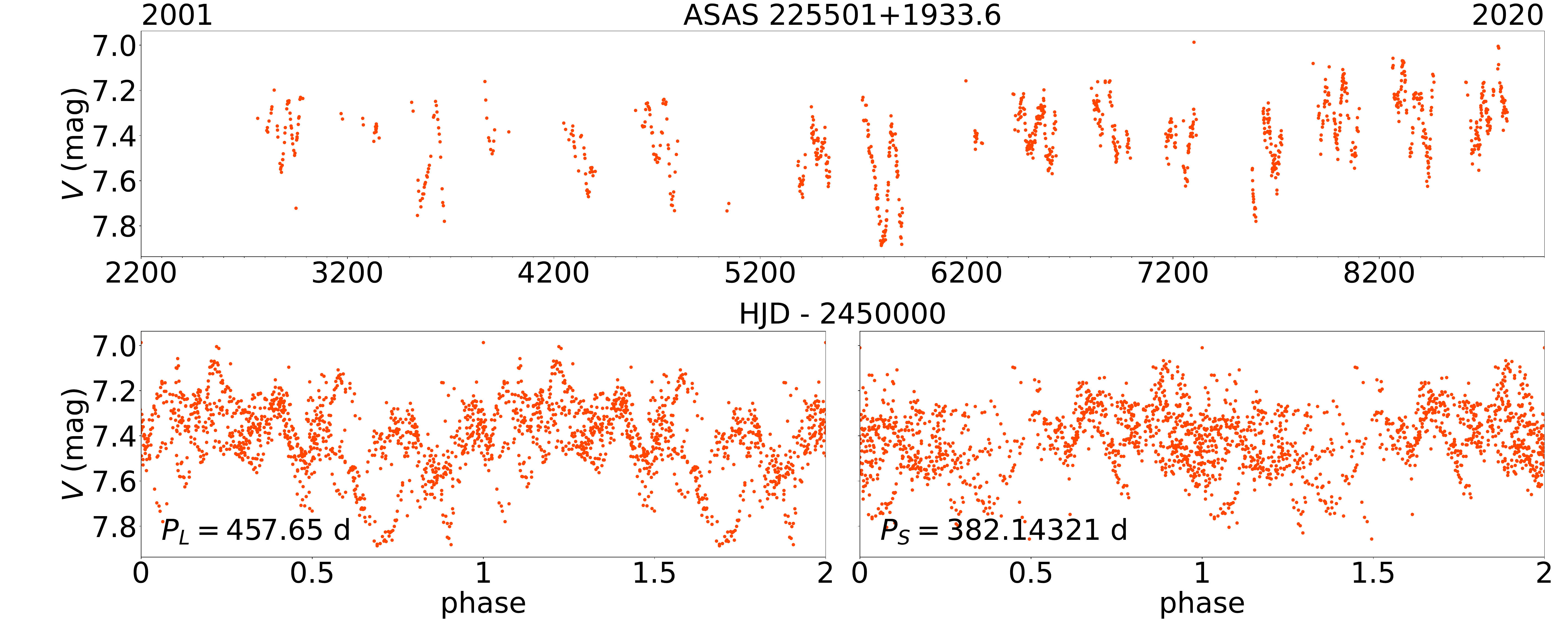}
\end{center}

\FigCap{Four examples of stars classified by \citet{percy2023} as LSP, for which LSP is not clearly visible in the ASAS data. Each object is shown in three different panels. \textit{Top panel}: the full unfolded ASAS \textit{V}-band light curve with the first and last year of the observations marked above the panel together with the star's ID. \textit{Bottom left panel}: phase-folded light curve with the longer period $P_L$ (computed in this work), indicated inside the plot. \textit{Bottom right panel}: phase-folded light curve with the shorter period $P_S$ (computed in this work), also indicated inside the plot.}
\label{fig:percy_not_lsp}
\end{figure}

In Figure \ref{fig:percy_not_lsp}, we present four examples of stars classified by \citet{percy2023} as LSP variables, namely ASAS~065821+0610.0 (RV~Mon), ASAS~070803-1155.4 (W~CMa), ASAS~202926+0953.9 (CT~Del), and ASAS~225501+1933.6 (GO~Peg). All these objects (including the remaining 15 stars classified as LSP by \citet{percy2023}) exhibit some form of long-term variability, however, based on the available data, it is difficult to unambiguously determine whether genuine LSPs are present within this sample. In particular, the difference between the amplitudes of the pulsations and the long-term variability is not sufficiently pronounced to allow a clear classification of these objects as LSP variables. The primary aim of this work was to compile a sample of unambiguous LSP stars that can be investigated further using complementary observational and theoretical methods. Consequently, for the 19 variables discussed here, we conclude that their long-term variability is most likely caused by dust ejections or multimode SRV behavior. Nevertheless, given the limitations of the current data, the presence of LSPs in these objects cannot be excluded.

\section{Conclusions} \label{sec:conclusions}

We carried out a systematic search for bright LSP stars using \textit{V}-band photometry from the ASAS. Starting from $16\,381$ bright and red Tycho-2 stars (\textit{V}$<8$~mag, $B-V>0.5$~mag, with limitation to the declination $<+28^\circ$), we analyzed $27\,632$ variables, determined their periods (long $P_L$ and pulsation $P_S$), and performed a~detailed visual classification. This procedure yielded a final sample of $23$ bright stars showing clear and robust signatures of LSP variability. Thirteen of them are new LSP discoveries, whereas $10$ were previously known as LSP stars and are independently confirmed here.

For all objects, we provided homogeneous ASAS photometry, refined values of $P_L$ and $P_S$, cross-identifications with Gaia DR3, 2MASS, VSX, SIMBAD, and distances with uncertainties from \citet{bailer-jones2021}. The stars from our sample are intrinsically bright and relatively nearby, making them prime targets for future high-resolution spectroscopy, radial velocity, and interferometric follow-up observations aimed at testing binary-dust scenario for the origin of LSPs. All ASAS light curves and Table \ref{tab:lsp} are released in a machine-readable form to facilitate such studies.

\section*{Acknowledgements}

We are indebted to the OGLE collaboration for the use of 
facilities of the Warsaw telescope at Las Campanas Observatory, Chile, for their permanent support and 
maintenance of the ASAS instrumentation, and to the Observatories of the Carnegie 
Institution of Washington for providing the excellent site for the observations.
PI and DMS acknowledge support from the European Union (ERC, LSP-MIST, 101040160). Views and opinions expressed are however those of the authors only and do not necessarily reflect those of the European Union or the European Research Council. Neither the European Union nor the granting authority can be held responsible for them. This work has been funded by the National Science Centre, Poland, through grant 2022/45/B/ST9/00243 to IS.

\bibliographystyle{acta}

\end{document}